\documentclass[aps,preprint,superscriptaddress,nofootinbib,preprintnumbers,eqsecnum,prd]{revtex4-1}

\usepackage{bm,hyperref,amsmath,amssymb,amsfonts,slashed,array,graphicx,soul,multirow,mathtools}

\newcommand{\twin}[1]{{#1^{\prime}}}
\newcommand{\LambdaQ}{\Lambda_\text{qcd}}

\graphicspath{}

\begin{document}

\title{Gamma-rays from Dark Showers with Twin Higgs Models}
\author{Marat Freytsis}
\affiliation{Institute of Theoretical Science, University of Oregon, Eugene, OR 97403}
\author{Simon Knapen}
\author{Dean J. Robinson}
\affiliation{Department of Physics, University of California, Berkeley, CA 94720, USA}
\affiliation{Ernest Orlando Lawrence Berkeley National Laboratory, University of California, Berkeley, CA 94720, USA}
\author{Yuhsin Tsai}
\affiliation{Maryland Center for Fundamental Physics, Department of Physics, University of Maryland, College Park, MD 20742, USA}
\begin{abstract}
We consider a twin WIMP scenario whose twin sector contains a full dark copy of the SM hadrons, where the lightest twin particles are twin pions. By analogy to the standard WIMP paradigm, the dark matter (DM) freezes out through twin electroweak interactions, and annihilates into a dark shower of light twin hadrons. These are either stable or decay predominantly to standard model (SM) photons. We show that this `hadrosymmetric' scenario can be consistent with all applicable astrophysical, cosmological and collider constraints. In order to decay the twin hadrons before the big-bang nucleosynthesis epoch, an additional portal between the SM and twin sector is required. In most cases we find this additional mediator is within reach of either the LHC or future intensity frontier experiments. Furthermore, we conduct simulations of the dark shower and consequent photon spectra. We find that fits of these spectra to the claimed galactic center gamma-ray excess seen by \emph{Fermi}-LAT non-trivially coincide with regions of parameter space that both successfully generate the observed DM abundance and exhibit minimal fine-tuning.
\end{abstract}
\maketitle

\tableofcontents
\clearpage 

\section{Introduction}
The twin Higgs (TH) mechanism provides a natural means to stabilize the electroweak scale up to $\mathcal{O}(10\,\textrm{TeV})$ with uncolored top partners, that belong to a twin dark copy of the standard model (SM) \cite{Chacko:2005pe,Chacko:2005un}. This framework thus can relax the evolving tension between naturalness expectations and the current limits on traditional top partners, and is therefore an attractive solution to the little hierarchy problem.

The central idea behind the TH mechanism is that the Higgs is realized as a pseudo-goldstone boson of an accidental global symmetry, which suffices to protect its mass from one loop quadratic corrections. In the simplest case, this structure is the result of an (approximate) $\mathbb{Z}_2$ exchange symmetry between the SM and twin sectors. (See Refs.~\cite{Craig:2014roa,Craig:2014aea} for a systematic discussion of more exotic options.)  In order for the $\mathbb{Z}_2$ exchange symmetry to provide sufficient protection \cite{Craig:2015pha}, the matter content of the twin sector must at the very least contain a twin third generation of quarks -- a twin top and twin bottom -- charged under twin color $SU(3)_{c}$ and twin electroweak $SU(2)_L$ gauge groups, such that the twin top yukawa and twin $SU(2)_L$ coupling are close to the corresponding SM values. Naturalness does not significantly constrain the remaining features of the twin sector, but all TH models contain a lightest twin particle (LTP) and a lightest twin hadron (LTH). The latter is either a bound state of twin quarks or a twin glueball. TH models moreover contain all the structure required to incorporate dark matter (DM) candidates, in the form of either a twin weakly charged WIMP \cite{Craig:2015xla,Garcia:2015loa} -- hereafter, a `T-WIMP' -- or twin baryonic asymmetric dark matter \cite{Farina:2015uea,Garcia:2015toa}. In this work we focus on TH models featuring T-WIMP candidates.

Although they require no additional light states with SM charges, twin Higgs models may nevertheless produce striking signatures in a variety of direct and indirect searches. The effectiveness of the various probes is mostly determined by the lifetime, decay modes and masses of the LTH and LTP, and as such it is useful to sketch out the space of signatures in terms of the properties of these two particles. In detail: (i) Relativistic LTPs at either the BBN or CMB epochs lead to strong tensions with current bounds on the effective SM neutrino degrees of freedom $\Delta N_{\mathrm{eff}}$; (ii) Twin states may be produced at the LHC through an exotic decay of the SM-like Higgs or its heavier partner, or through the decay of some of the TeV-scale, colored states that could be accessible in certain UV completions. In some TH models, this can lead to interesting LHC signatures from the decay of the LTH \cite{Craig:2015pha,Curtin:2015fna,Csaki:2015fba,Cheng:2015buv}; (iii) A LTH that is metastable on BBN timescales may lead to either overclosure, arising from other (meta)stable hadrons with masses nearby to the LTH, or post-BBN matter-dominated eras that disrupt the standard BBN paradigm; (iv) Annihilation of T-WIMP candidates into hard $b$ quarks, charm quarks or $\tau$'s may create tensions with evolving astrophysical bounds on antiproton or positron production in the galactic central stellar cluster (see e.g. \cite{Bergstrom:2013jra,Bringmann:2014lpa,Bringmann:2009ca,Cirelli:2014lwa,Pettorino:2014sua,Cholis:2014fja}). Finally, apart from these LTH and LTP signatures, one may also search for deviations of Higgs couplings. These are in principle the most robust probes of twin Higgs models, but their reach at the LHC is rather limited \cite{Burdman:2014zta}. 

Several TH models have been previously proposed, that may be broadly characterized by these signatures. The first instance is the `mirror twin Higgs' \cite{Chacko:2005pe}, which includes a full copy of the SM in the twin sector. The LTPs are therefore twin neutrinos and twin photons, while the LTH is a twin pion that rapidly decays to twin leptons and twin photons well before BBN. This model is susceptible to strong BBN and CMB bounds on $\Delta N_{\mathrm{eff}}$. To evade these, one either requires an asymmetric reheating between both sectors \cite{Berezhiani:1995am,Foot:2004pa,Berezhiani:2005ek} or require large entropy production during the QCD phase transition \cite{Chacko:2005pe,Barbieri:2005ri,Chang:2006ra}. 

Two other recent TH proposals include the `fraternal twin Higgs' \cite{Craig:2015pha} and the `vector-like twin Higgs' \cite{vectortwin}. In these constructions, the low energy spectrum of the twin hadronic sector is given by a zero or one light flavor twin QCD theory. The LTH is then either a glueball or an onium state whose mass is determined by the twin confinement scale,  $\twin{\LambdaQ}$, that is itself constrained to be nearby the SM confinement scale, $\LambdaQ$. In the vector-like twin Higgs model, the LTH and the LTP are the same particle, but in the fraternal twin Higgs model the LTP can be either a twin neutrino or the LTH itself. The decays of the LTH may generate (possibly displaced) collider signatures, while a twin neutrino LTP contributes $\Delta N_{\mathrm{eff}}\gtrsim 0.075$ \cite{Craig:2015xla,Garcia:2015loa}, that is potentially detectable in the future.  These models can further admit a twin tau T-WIMP that annihilates into twin hadrons, some of which in turn decay to $b\bar{b}$ or $\tau\bar{\tau}$ pairs (see e.g. \cite{Garcia:2015loa}), potentially detectable in (future) astrophysical data. 

\begin{table}[t]
	\renewcommand*{\arraystretch}{1.25}
	\newcolumntype{C}{ >{\centering\arraybackslash} m{1.2cm} <{}}
	\newcolumntype{D}{ >{\scriptsize\centering\arraybackslash} m{2cm} <{}}
	\newcolumntype{E}{ >{\scriptsize\raggedleft\arraybackslash} m{2.4cm} <{}}
	\begin{center}
		\begin{tabular*}{0.8\linewidth}{@{\extracolsep{\fill}}C||ED|ED}
			\hline
			& \multicolumn{2}{c|}{\textbf{Hadrosymmetric}} & \multicolumn{2}{c}{\textbf{Vector-like/Fraternal}}\\[8pt]
			&	LTP:						&$\twin{\pi}$							&LTP:						&onium/glueball\\
			&	LTH:						&$\twin{\pi}$							&LTH:						&onium/glueball\\
			&	$\Delta N_{\textrm{eff}}$:		&$0$									&$\Delta N_{\textrm{eff}}$:		&$0$\\
			\multirow{-5}{*}{\rotatebox[origin=c]{90}{\begin{tabular}{c}$m_{\textrm{LTP}} > T_{\textrm{BBN}}$\\{\scriptsize $\Delta N_{\textrm{eff}} =0$}\end{tabular}}}
			&	T-WIMP ann: 				&$\to \#(\gamma\gamma)$, $\#(\ell\bar{\ell})$ 	&T-WIMP ann: 					&$\to \#(b\bar{b})$, $\#(\tau\bar{\tau})$\\
			\hline
			& \multicolumn{2}{c|}{\textbf{Mirror}} & \multicolumn{2}{c}{\textbf{Fraternal}}\\[8pt]
			&	LTP:						&$\twin{\gamma}$, $\twin{\nu_i}$		&LTP:						&$\twin{\nu}$\\
			&	LTH:						&$\twin{\pi}$		&LTH:						&onium/glueball\\
			&	$\Delta N_{\textrm{eff}}$: 		&excluded							&$\Delta N_{\textrm{eff}}$:		&0.075\\
			\multirow{-5}{*}{\rotatebox[origin=c]{90}{\begin{tabular}{c}$m_{\textrm{LTP}} < T_{\textrm{BBN}}$\\{\scriptsize $\Delta N_{\textrm{eff}} >0$}\end{tabular}}} 
			&	T-WIMP ann: 				&no T-WIMP						&T-WIMP ann: 					&$\to \#(b\bar{b})$, $\#(\tau\bar{\tau})$\\
			\hline\hline
			& \multicolumn{2}{c|}{\begin{tabular}{c}$m_{\textrm{LTH}} < \twin{\LambdaQ}$\\{\scriptsize no visible LTH decays at LHC}\end{tabular} }& \multicolumn{2}{c}{\begin{tabular}{c}$m_{\textrm{LTH}} > \twin{\LambdaQ}$\\{\scriptsize possible (displaced) LTH decays at LHC}\end{tabular}}
		\end{tabular*}
	\end{center}
	\caption{Schematic overview of theory space of twin Higgs models, as classified by various signatures and features. Twin states are denoted by a prime.}
	\label{tab:introsum}
\end{table}

In Table I we schematically summarize the signature space of these twin Higgs models according to the signatures and potential features (i)-(iv) above. Laid out in this fashion, it becomes clear that there is a fourth category of TH models, which is so far largely unexplored in the Literature. Namely, the case where the LTP is heavy enough to trivially avoid $\Delta N_{\mathrm{eff}}$ constraints, while the LTH is at the same time always light enough, and therefore metastable enough, to be invisible at the LHC. While certain parts of the parameter space of the fraternal and vector-like twins realize this scenario, we consider here models in which these features are robust predictions everywhere in parameter space. A well-motivated and representative example is a twin sector that contains a mirror copy of the SM hadrons -- i.e. all three twin quark generations with three light quark flavors -- but no dark radiation -- i.e. no light leptons or photons. A T-WIMP may be incorporated as an additional Dirac or Majorana state. We call this scenario the \emph{`hadrosymmetric twin Higgs'}. 

In this scenario, the existence of multiple light twin quark flavors produces a twin pion LTH, a pseudo-Nambu-Goldstone boson (pNGB). Unlike in zero and one light flavor twin sectors, here the pion mass may be much lighter than the twin confinement scale. The absence of dark radiation ensures the twin pion is also the LTP and avoids, in the most straightforward manner, tension with $\Delta N_{\mathrm{eff}}$ bounds, provided the twin pion is heavier than $T_{\rm BBN} \sim$ MeV.  Without dark radiation decay modes, the twin pion must promptly decay by BBN to SM degrees of freedom to avoid overclosure or a matter-dominated era thereafter by twin pNGBs.  On collider timescales the twin pion is then stable (cf. e.g. Refs \cite{Schwaller:2015gea,Cohen:2015toa}) and appears as missing energy produced with only a small cross-section: From the viewpoint of collider and $\Delta N_{\mathrm{eff}}$ searches, the hadrosymmetric scenario then represents a `least detectable' scenario compared to the other models in Table I. 

We show in this work, however, that these generic features have several implications, that, along with the BBN lifetime bound, provide other means to effectively probe the hadrosymmetric scenario. First, since the LTP is a pNGB and therefore a pseudoscalar, it cannot decay efficiently to SM states through the Higgs portal. A sufficiently prompt twin pion decay by BBN therefore requires the existence of an additional twin-SM mediation scale. (This situation is similar to the case with one light flavor, where the LTH is $0^{-+}$ meson \cite{Garcia:2015loa,Farina:2015uea}.)  There are a limited number of choices for such a portal, especially if one wants to retain a minimal twin-Higgs sector and not introduce additional SM-charged particles. There are two phenomenologically different regimes for this portal, corresponding to whether the twin pion mass is above or below three times the SM pion mass, $3m_\pi$: Since the twin pNGBs can be much lighter than the twin confinement scale, the twin pion mass is a free parameter of the theory. In the case that the twin pion mass is above $3m_\pi$, we find the portal does not require a UV completion below the twin Higgs cutoff scale. However, if the twin pion mass is below $3m_\pi$, we show, for a representative mediation model, that in most of the viable parameter space the BBN lifetime bound implies that the mediator itself should be accessible either at the LHC or at the intensity frontier.

Second, the strongly-coupled twin sector generically acts like a neutral natural hidden valley \cite{Strassler:2006im,Han:2007ae}, and in the spirit of Ref.~\cite{Freytsis:2014sua}, DM annihilations into twin quarks can therefore produce `dark showers' of twin hadrons, that then subsequently decay to the SM sector. In the case that the twin pions are lighter than $3m_\pi$, a leptophobic mediator ensures that twin pion decays are predominantly to diphotons. If instead they are heavier than $3m_\pi$, the twin pions decay mainly to three SM pions and hence to a higher multiplicity of softer photons or leptons. The indirect detection signatures for dark showers are then mainly encoded into photons, as in Ref.~\cite{Freytsis:2014sua}, evading present or future astrophysical cosmic-ray bounds on the associated emission of antiprotons and positrons. 

This dark showering to photons may produce a visible astrophysical signal. We present detailed numerical simulations of the photon spectra produced by Dirac and Majorana T-WIMP annihilation in the galactic center, compared against $\gamma$-ray data from \emph{Fermi}-LAT. We find that for a Dirac T-WIMP, there is a viable region of parameter space in which these dark showers may reproduce the putative galactic center $\gamma$-ray excess (GCE), claimed to be seen in \emph{Fermi}-LAT data from the central regions of the Milky Way galaxy~\cite{Goodenough:2009gk,Hooper:2010mq,Hooper:2011og,Hooper:2013rwa,Abazajian:2014fta,Daylan:2014rsa,Calore:2014nla,Calore:2014xka,TheFermi-LAT:2015kwa} and dwarf spheroidal galaxies~\cite{Geringer-Sameth:2015lua,Hooper:2015ula}, as well as simultaneously generate an appropriate amount of DM and exhibit minimal fine-tuning. (See also Refs.~\cite{TheFermi-LAT:2015kwa, Abazajian:2012dgr,Abazajian:2012edg,Gordon:2013vta,Cholis:2014lta,Cholis:2014noa,Lee:2014mza,Lee:2015fea} which discuss alternative astrophysical explanations for the GCE, or dispute observations of an excess in dwarf spheroidal galaxies \cite{Drlica-Wagner:2015xua}.) Alternatively, treating observed photon fluxes as a conservative upper bound,  there are large regions of parameter space are not excluded by current galactic center $\gamma$-ray data.

This paper is organized as follows. In Sec.~\ref{sec:twin} we present the hadrosymmetric twin Higgs model and benchmark dark matter models, followed by an examination of all applicable astrophysical, cosmological and collider bounds in Sec.~\ref{sec:bounds}. In Sec.~\ref{sec:TPDM} we consider effective field theory analyses for portals mediating the twin pion decay, as well as a sample UV completion and corresponding constraints on its mediator. Simulations of dark showering DM annihilations and corresponding photon spectra are presented in Sec.~\ref{sec:indirect}, along with both fits to GCE spectra and constraints from $\gamma$-ray fluxes.

\section{Twin sector\label{sec:twin}}
\subsection{Twin Higgs framework}
In the twin Higgs framework the SM Higgs is realized as the pNGB of a spontaneously broken global symmetry. The key difference with traditional pNGB Higgs frameworks is that the symmetry in question is \emph{accidental} rather than exact. This has the advantage that the top partner may be fully neutral under the SM gauge groups, which removes conventional collider constraints on the top partner. On the other hand, this also implies that the accidental symmetry is generally not radiatively stable beyond 5 to 10 TeV, at which scale a UV completion is then required. The known options for UV completions are supersymmetry \cite{Chang:2006ra,Craig:2013fga,Falkowski:2006qq}, compositeness \cite{Batra:2008jy,Barbieri:2015lqa,Low:2015nqa}, orbifolds \cite{Craig:2014roa,Craig:2014aea} and holographic setups \cite{Geller:2014kta,Csaki:2015gfd}.

The original version of the TH model \cite{Chacko:2005pe} consists of a complete dark copy of the SM gauge and matter content, which are mapped into each other by $\mathbb{Z}_2$ symmetry. The Higgs potential is further assumed to have an approximate $SU(4)$ symmetry, of which the SM Higgs, $H$, and twin Higgs, $\twin{H}$, furnish a $SU(4)$ fundamental $(H,\twin{H})$.\footnote{For strongly coupled UV completions an explicit $SO(8)$ \cite{Barbieri:2005ri,Chacko:2005un} or $Sp(4)\times Sp(4)$ group \cite{Batra:2008jy} is needed to ensure that custodial symmetry is preserved. In this paper we are agnostic about the specific UV completion and we therefore do not make this distinction here.} (Hereafter, we denote all twin sector objects by a prime.) In the broken phase of both the SM and twin $SU(2)_L$, the remaining physical degrees of freedom in the Higgs sector are a pNGB, $h$, and a heavy radial mode. The pNGB $h$ is then naturally light, and is identified with the scalar particle observed at the LHC. 

Even if the accidental $SU(4)$ symmetry holds at tree-level, it is broken at one loop by the presence of Yukawa and gauge couplings. As such, $h$ will then receive important radiative corrections. The key feature of the TH framework is, however, that the most dangerous correction -- the one-loop quadratic divergence -- is cancelled by the presence of the $\mathbb{Z}_2$ exchange symmetry between the SM and twin sectors. 

In many models, the $\mathbb{Z}_2$ is itself only an approximate symmetry, mainly to avoid introducing too many extra light degrees of freedom in tension with $\Delta N_{\textrm{eff}}$ bounds. The quality of the $\mathbb{Z}_2$ that is needed to adequately cancel the quadratic divergence depends on the cut-off, $\Lambda$, of the Twin Higgs setup. For $\Lambda \sim 5$ TeV, one requires \cite{Craig:2015pha}
\begin{equation}
\label{eqn:TCD}
\frac{|y_t - y_t' |}{y_t}\sim 0.01\,,\qquad
\frac{|g_2 - g_2' |}{g_2}\sim 0.1\,,\qquad\text{and}\qquad
\frac{|g_3 - g_3' |}{g_3}\sim 0.1\,.
\end{equation}
where $y_t$ ($\twin{y_t}$) is the top (twin top) yukawa, and $g_3$ ($\twin{g_3}$) is the (twin) color gauge coupling. There are no restrictions on the twin hypercharge coupling, $\twin{g_1}$, nor on the remaining twin yukawa couplings, other than that they cannot be $\mathcal{O}(1)$. The complete absence of the first two generations, or a gauged twin hypercharge, do not reintroduce an appreciable amount of fine-tuning. In this paper we consider the `hadrosymmetric' twin Higgs scenario, for which the twin sector contains twin copies for all three generations of quarks, but the twin lepton sector is absent. We also assume that twin hypercharge, if present, is at most only a global symmetry, but is not gauged.

As in traditional pNGB Higgs models, the twin Higgs setup generically predicts mixing between the SM and twin Higgs, that results in $\mathcal{O}(1)$ Higgs coupling deviations. To reduce this mixing to acceptable levels, the SM sector vacuum expectation value $v\equiv \langle H\rangle$ must be tuned down with respect to the vacuum expectation in the twin sector $f\equiv \langle H'\rangle$. To be consistent with the current status of the Higgs branching ratio fits, one requires $f/v \gtrsim 3$ (see Ref.~\cite{Craig:2015pha,Burdman:2014zta} and Sec.~\ref{sec:HIW} below), which corresponds to $\mathcal{O}(20\%)$ fine-tuning. (See, however, \cite{Beauchesne:2015lva} for a recent example with a smaller fine-tuning.) In principle, there may be several ways for the SM to communicate with the twin sector, but the twin-Higgs mixing with the SM Higgs is the only portal common to all models.

In the absence of gauged twin hypercharge, the heavy $\twin{SU(2)_L}$ gauge bosons are degenerate, $m_{\twin{Z},\twin{W}} = \twin{g_2} f/2$, so that their masses scale with a factor of $f/v$ with respect to their SM counterparts.   Na\"{i}vely, all twin quarks are similarly typically $f/v$ times heavier than their SM counterparts. However since all twin yukawas except $\twin{y_t}$ are almost unconstrained, this assumption may be relaxed for all quarks except for the twin top. For most of the discussion in this paper, we nevertheless assume the twin yukawas are set by the SM-twin $\mathbb{Z}_2$ such that $\twin{y} = y$, and the twin quark masses are then
\begin{equation}
\label{eqn:TQM}
m_{\twin{q}} =  f/v\,m_{q}\,,
\end{equation}	
for all $\twin{q}$.

\subsection{Twin hadron spectrum}
For the  $\mathcal{O}(10\%)$ deviation of twin QCD coupling, $\twin{g_3}$, permitted in eq.~\eqref{eqn:TCD}, computing the two-loop renormalization group evolution from the UV cutoff down, one finds a corresponding $\mathcal{O}(1)$ deviation in the twin confinement scale compared to the SM, viz.
\begin{equation}
\label{eqn:LQR}
\lambda \equiv \twin{\LambdaQ}/ \LambdaQ \sim 0.2\, \mbox{--}\, 5\,.
\end{equation}
With two or three light flavors, the dependence of $\lambda$ on the mass spectrum of the twin fermions is rather mild. We shall therefore treat the confinement scale ratio $\lambda$ as a free parameter of the twin theory, within the range indicated by eq.~\eqref{eqn:LQR}.

We characterize twin quarks as light or heavy depending on whether their mass is small or large compared to the scale of chiral symmetry breaking or to the confinement scale, as appropriate. Hadronic matrix elements of na\"{i}ve mass dimension $p$ typically scale as $\lambda^p$. As an immediate example, the masses of non-pNGB light quark hadrons typically scale as 
\begin{equation}
\label{eqn:THM}
m_{\twin{\textrm{had}}} \simeq m_{\textrm{had}} \lambda\,,
\end{equation}
and the twin pion decay constant $f_{\twin{\pi}} \simeq f_{\pi} \lambda$. (We adopt the convention under which $f_\pi \simeq 93$~MeV.) The masses of heavy quark mesons, in contrast, scale linearly with the heavy quark mass.

\begin{figure}[tb]
	\centering\includegraphics[width=8cm]{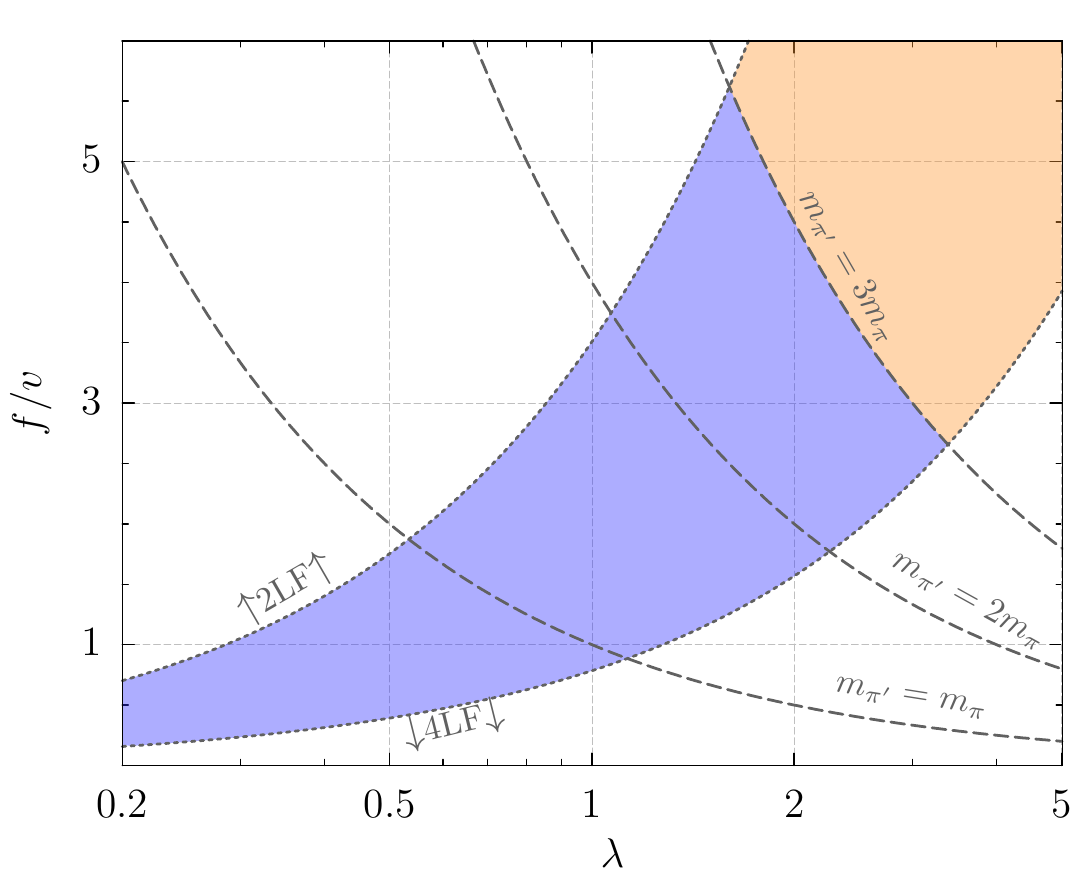}
	\caption{
		Approximate region for $\lambda$ and $f/v$ for a three light flavor theory (blue and orange). Empirical bounds on the parameter space are not shown (see Secs.~\ref{sec:bounds} and \ref{sec:indirect} below). Also shown are contours of constant $m_{\twin{\pi}}/m_{\pi}= 1$, $2$, and $3$ (dotted gray). In the blue (orange) region, twin pion decays are predominantly to diphotons (pions); see Sec.~\ref{sec:TPDM} below.}
	\label{fig:LR}
\end{figure}

We consider twin sectors with at least two light twin quark flavors, so that the hadronic spectrum contains pNGBs. We focus mainly on the case that the twin sector has exactly three light twin quarks, $\twin{s}$, $\twin{d}$ and $\twin{u}$, just as on the SM side. To be certain that the strange is light while the charm is heavy, one requires $m_{\twin{s}} \ll 4\pi f_{\twin{\pi}} \sim \lambda \times 1$~GeV, the approximate scale of chiral symmetry breaking, while $m_{\twin{c}} > 4\pi f_{\twin{\pi}}$. Combining with eq.~\eqref{eqn:TQM}, this implies the constraints
\begin{equation}
\label{eqn:LQCDB}
\lambda  \gg f/v \frac{m_s}{4\pi f_\pi} \qquad  \mbox{and}  \qquad   \lambda   <  f/v \frac{m_c}{4\pi f_\pi}\,. 
\end{equation}
From eqs.~\eqref{eqn:TQM} and \eqref{eqn:THM} one further expects
\begin{equation}
\label{eqn:TPM}
m_{\twin{\textrm{pNGB}}} \simeq m_{\textrm{pNGB}} \sqrt{\lambda f/v }\,.
\end{equation}
Combining eqs. \eqref{eqn:LQR} and \eqref{eqn:LQCDB} with eq.~\eqref{eqn:TPM}, one may hence determine the approximate allowed $f/v$--$\lambda$ parameter space for a three light flavor theory, as well as contours of constant $m_{\twin{\textrm{pNGB}}}/m_{\textrm{pNGB}}$. In Fig.~\ref{fig:LR} we show this approximate region.

As in the SM sector, the lightest pNGBs, and thus the lightest twin hadrons, are twin pions, followed by heavier twin kaons and $\eta$'s. The twin pions embed into a twin isospin triplet ($\pi'^\pm,\pi'^0$). Unlike in the SM sector, in the absence of a gauged twin hypercharge, twin isospin breaking arises only from the splitting of the yukawas. In particular, twin pion mass splitting effects arise only at second order in isospin breaking, such that
\begin{equation}
\frac{m^2_{\twin{\pi}^+} - m^2_{\twin{\pi}^0}}{(4\pi f_{\twin{\pi}})^2} \sim \delta^2\,, \qquad \delta \sim \frac{ m_{\twin{d}} - m_{\twin{u}}}{4\pi f_{\twin{\pi}}} =  \frac{ f/v}{\lambda} \delta_{\textrm{SM}}\,,
\end{equation}
in which $\delta_{\textrm{SM}} \sim 1\%$ is the isospin breaking in the SM sector induced by splitting of the up-down yukawas. In most of the parameter space, we therefore expect the mass splitting of the $\twin{\pi}^0$ and $\twin{\pi}^\pm$ to be much smaller than for the SM pions, but with the same sign.  

In the class of twin sector scenarios we consider in this paper, the twin pions $\twin{\pi}^{\pm}$ are charged under a (residual) global $U(1)$ and therefore stable. In the absence of a light twin photon or other twin-SM mediator, the $\twin{\pi}^0$ can only decay via the Higgs portal to SM final states. This decay requires both parity and twin isospin violation and is therefore heavily suppressed. We discuss this and twin-SM mediator models for decay of the twin pion in further detail in Sec.~\ref{sec:TPDM}. 

Apart from pions, the remainder of the twin light hadron spectrum comprises: higher spin states, such as the $\twin{\rho}$ or $\twin{\omega}$; a twin $\eta'$; glueballs such as the $\twin{f_0}$; and baryons, such as the twin proton $\twin{p}$ and neutron $\twin{n}$. Similar expectations apply to the expected twin proton-neutron mass splitting, as for the pion mass splitting. The twin proton is stabilized by baryon number, while the heavier twin neutron is stable as well due to the absence of light twin leptons.  We assume all heavier hadrons strongly or weakly decay to pions, $\twin{n}$ and $\twin{p}$, as in the SM sector, and that there is no twin baryon asymmetry. 

\subsection{Dark matter}
We consider two different twin sector benchmark scenarios that produce respectively Majorana or Dirac DM, coupled to the twin sector by twin electroweak interactions. Along with three generations of twin quarks -- electroweak doublets $\twin{Q}^i$ and singlets $\twin{u}^{c\,i}$ and $\twin{d}^{c\,i}$, for $i =1,2,3$ -- we restore a single lepton-like generation -- an electroweak doublet $\twin{L}$ and a singlet $\twin{E}$ -- and an anomalous global twin hypercharge symmetry. This single electroweak doublet is the minimal content required to cancel the $\twin{SU(2)_L}$ Witten anomaly. One then obtains an effective theory 
\begin{equation}
\label{eqn:DMHC}
y_E \twin{H}^\dagger \twin{L}^c \twin{E} + \frac{1}{\Lambda} \twin{H} \twin{L}^c \twin{H} \twin{L}^c \,,
\end{equation}
leading to one Dirac and one Majorana state. (As an alternative, one might consider a $\twin{SU(2)_L}$ triplet DM candidate, which directly acquires a Majorana mass term $m\chi\chi$.) 

We hereafter denote the Dirac and Majorana states respectively by $\psi$ and $\chi$, with mass $m_\psi=y_E f/\sqrt{2}$ and $m_{\chi}=f^2/\Lambda$. Twin hypercharge aside, both states are stabilized by fermion number. (Note there is also an accidental twin baryon number in the twin quark sector.) In the Majorana (Dirac) DM scenario we assume $m_\chi < m_\psi$ ($m_\psi < m_\chi$), and assume further $m_\psi$ ($m_\chi$) is sufficiently heavy that its contribution to the DM relic density can be neglected.  The effective theory \eqref{eqn:DMHC} should only be considered a representative of models containing either Dirac and Majorana DM with Higgs couplings of the form in eq.~\eqref{eqn:DMHC}, that may be UV completed without anomalous global symmetries. 

\section{Cosmological and terrestrial bounds}
\label{sec:bounds}
\subsection{Relic density}

The thermally averaged $2 \to 2$ annihilation cross-section for fermions of mass $m_{\textrm{DM}}$ at temperature $T(x) = m_{\textrm{DM}}/x \ll m_{\textrm{DM}}$ has general form
\begin{equation}
\langle \sigma v \rangle = \frac{x^2 e^{2x}}{16 \pi m_{\textrm{DM}}^4 \mathcal{S}} \int_{4 m_{\textrm{DM}}^2}^\infty \!\!\!ds\, \Sigma(s)\sqrt{\frac{s}{4m_{\textrm{DM}}^2} -1} \,\, K_1\!\big(\sqrt{s}/T\big)\,.
\end{equation}
Here $\mathcal{S}$ is a symmetry factor -- $\mathcal{S} = 2$ for Majorana $\chi$ and $\mathcal{S} = 1$ for Dirac $\psi$ -- and $\Sigma(s)$ is the spin-summed square amplitude for annihilation to two (twin) quarks, integrated over the final state phase space. Including both annihilation to twin quarks through the twin $\twin{Z}$ as well as annihilation to SM $b\bar{b}$ via the Higgs portal, one finds 
\begin{align}
\Sigma_\chi(s) & \simeq \frac{3}{\pi f^4} \bigg\{ \frac{5}{3} \frac{s(s - 4m_{\chi}^2)}{(1 - s/m_{\twin{Z}}^2)^2} + \frac{8m_{\chi}^2m_b^2(s-4m_{\chi}^2)}{\big|s - m_h^2 + i m_h \Gamma_h\big|^2} \bigg\}\notag\\
\Sigma_\psi(s) & \simeq \frac{3}{\pi f^4} \bigg\{ \frac{5}{6} \frac{s(s - m_{\psi}^2)}{(1 - s/m_{\twin{Z}}^2)^2}  + \frac{2m_{\psi}^2m_b^2(s-4m_{\psi}^2)}{\big|s - m_h^2 + i m_h \Gamma_h\big|^2} \bigg\}\,,
\end{align}
where $m_{\twin{Z}} = m_Z f/v$ is the twin $Z$ mass and we treat the twin quarks as massless, so that annihilation through the twin $\twin{Z}$ proceeds to $N_f = 5$ twin quark flavors. The latter assumption, which greatly simplifies the square amplitude expressions, anticipates the result that $m_{\chi,\psi} \gg m_{\twin{b}}$ for $\Omega_{\chi,\psi} = \Omega_{\text{DM}}$; the regime that $m_{\chi,\psi} < m_{\twin{b}}$ cannot achieve a sufficient DM relic abundance. Corrections to $\langle \sigma v \rangle$ on the $\Omega_{\chi,\psi} = \Omega_{\text{DM}}$ contour arising from the $\twin{b}$ mass arise at the $\mathcal{O}(\%)$ level, and can be safely neglected. From the common $s - 4m_\chi^2$ factors, observe that the Majorana cross-section is $p$-wave suppressed, as expected from Fermi statistics. Comparing to the fraternal twin Higgs, the annihilation cross section is somewhat enhanced here because of the larger number ($N_f = 5$) of light twin quarks available as final states. The required DM mass will therefore be somewhat smaller than was obtained for the fraternal twin WIMPs \cite{Craig:2015xla,Garcia:2015loa}.

\begin{figure}[t]
	\includegraphics[width=7cm]{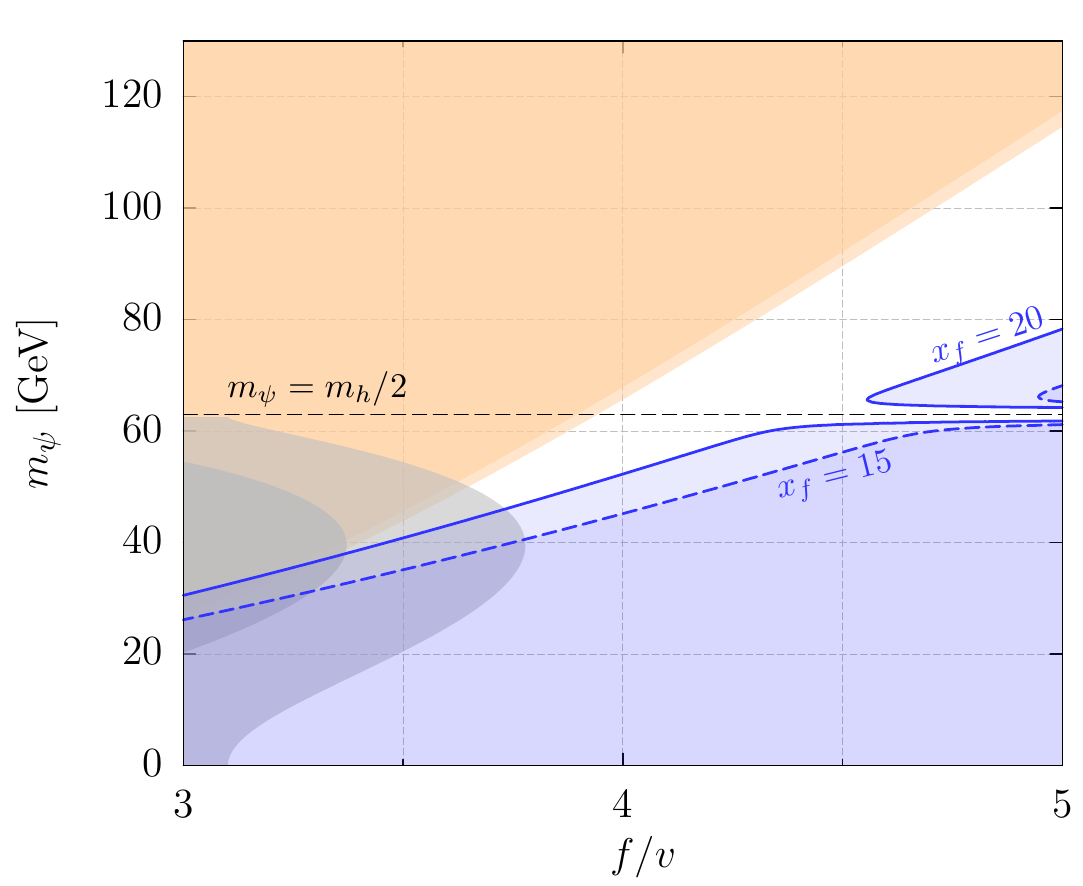}
	\hspace{1cm}
	\includegraphics[width=7cm]{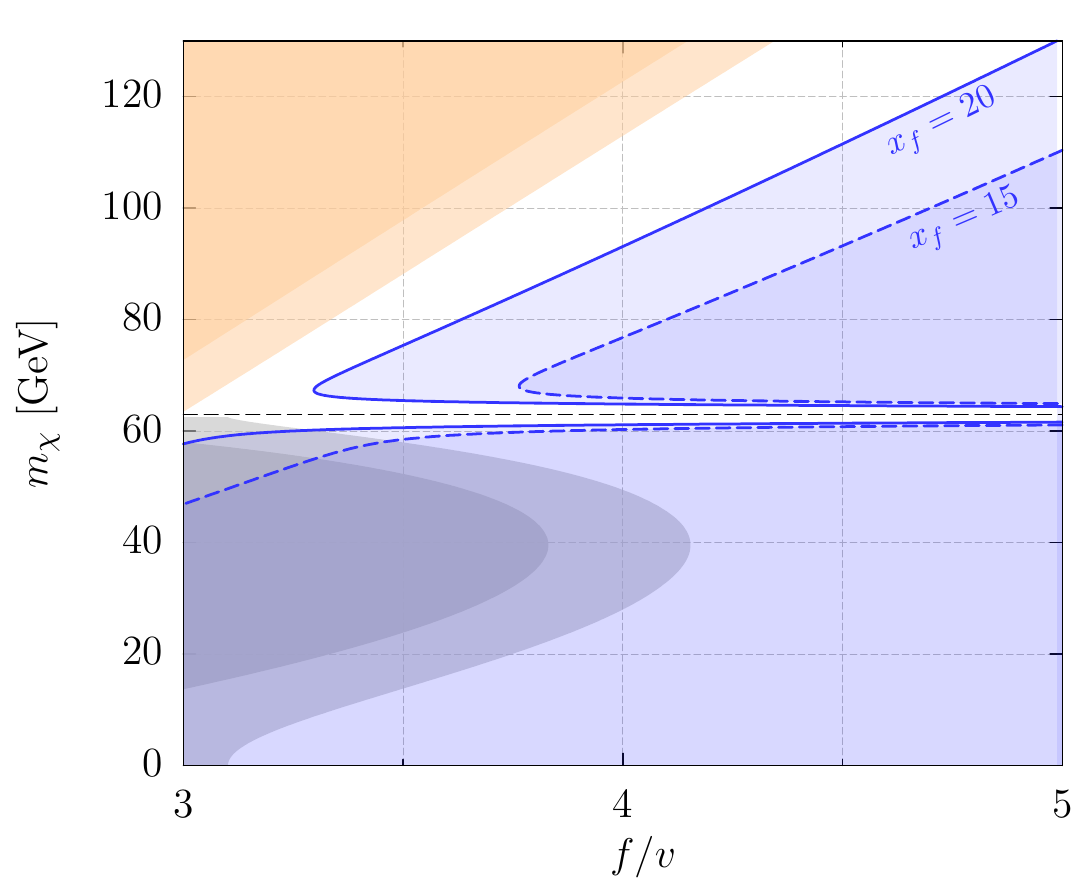}
	\caption{Exclusion regions for DM overproduction (blue), and for the Higgs invisible width including Higgs decays to both DM-DM and $\twin{b}\twin{\bar{b}}$ (light grey), in the Dirac (left) and Majorana (right) scenarios. The blue solid (dashed) line corresponds to a $100\%$ DM contribution for freeze-out temperature $x_f = 20$ ($x_f = 15$).  The narrow Higgs resonance induces a runaway in $f/v$ very close to $m_{\textrm{DM}} = m_h/2$. Also shown is the Higgs invisible width exclusion region for Higgs decays to DM-DM only (dark grey), and CMB reionization bounds for $x_f=20$ (dark orange) and $x_f = 15$ (light orange), assuming a maximal energy deposition fraction $f_{\textrm{eff}} = 1$. }
	\label{fig:RI}
\end{figure}

To a good approximation, one finds that the thermal relic abundances
\begin{subequations}
	\label{eqn:FVDM}
	\begin{align}
	\Omega_\chi h^2 & \simeq 0.12 \bigg[\frac{g_\chi(\mu^2,20/x_f)}{g_\chi(0.19,1)}\bigg]  \bigg[\frac{x_f}{20}\bigg]^2 \bigg[\frac{f/v}{3}\bigg]^4 \bigg[ \frac{59~\mbox{GeV}}{m_\chi} \bigg]^2\,,\\
	\Omega_\psi h^2 & \simeq 0.12 \bigg[\frac{g_\psi(\mu^2,20/x_f)}{g_\psi(0.05,1)}\bigg]\bigg[\frac{x_f}{20}\bigg] \bigg[\frac{f/v}{3}\bigg]^4 \bigg[ \frac{31~\mbox{GeV}}{m_\psi} \bigg]^2\,,
	\end{align}
\end{subequations}
in which the Higgs resonance contributions are not displayed; $x_f = m_{\textrm{DM}}/T_f$, the freeze-out temperature; the functions
\begin{subequations}
	\begin{align}
	g_\chi(\mu^2,20/x_f) & =  \frac{1}{320 + 43[20/x_f]} - 2\mu^2\frac{320 + 83[20/x_f]}{(320 + 43[20/x_f])^2} \,,\\
	g_\psi(\mu^2,20/x_f) & =  \frac{1}{320 + 59[20/x_f]} - 2\mu^2\frac{320 + 83[20/x_f]}{(320 + 59[20/x_f])^2}\,;
	\end{align}
\end{subequations}
and the parameter
\begin{equation}
\mu^2 = \frac{4 m_{\textrm{DM}}^2}{(m_Z f/v)^2} = 0.19 \bigg[\frac{m_\chi}{59~\mbox{GeV}}\bigg]^2\bigg[\frac{3}{f/v}\bigg]^2 =  0.05 \bigg[\frac{m_\psi}{31~\mbox{GeV}}\bigg]^2\bigg[\frac{3}{f/v}\bigg]^2\,,
\end{equation}
encodes corrections from the twin $\twin{Z}$ resonance.  For the Dirac case, the latter corrections can be safely neglected. Eq.~\eqref{eqn:FVDM} parametrizes the DM mass in terms of the order parameter ratio $f/v$. For the Majorana case, note that the required DM mass is in good agreement with the na\"\i ve scale of the mass terms generated by the operators \eqref{eqn:DMHC}. In Fig.~\ref{fig:RI} we show contours of $100$\% DM thermal production -- i.e. $\Omega_{\psi,\,\chi} = \Omega_{\textrm{DM}}$ -- for both $x_f = 20$ and $15$, including Higgs resonance contributions.

\subsection{Higgs invisible width}
\label{sec:HIW}
Higgs-twin Higgs mixing induces the operators $(v m_\chi/f^2) h \chi \chi$ and $(v m_\psi/f^2) h \bar{\psi}\psi$ respectively for Majorana and Dirac scenarios, which contribute to the Higgs invisible width. 
Current bounds on the Higgs branching ratio to invisible states have been extracted by a profile likelihood analysis of Higgs coupling measurements for the model-specific case of twin Higgs \cite{Craig:2015pha,Craig:2015xla}, and are sensitive to $f/v$. In Fig.~\ref{fig:RI} we show corresponding exclusion regions for both scenarios, to be compared to the DM production contours \eqref{eqn:FVDM}. The large Majorana DM masses -- typically $m_\chi > m_h/2$ -- automatically renders the Higgs invisible width bound inapplicable to this scenario.

If $y_{\twin{b}} = y_b$, as we generally assume, then a sizable $h \to \twin{b}\twin{\bar{b}}$ contribution requires $f/v \gtrsim 4$ in order for the Dirac $\psi$ to both generate a sufficient DM abundance and be consistent with current Higgs to invisible branching ratio data. We may relax this bound, however, by letting the twin $\twin{b}$ mass float down so that contributions to the Higgs invisible width from $h \to \twin{b}\twin{\bar{b}}$ may be neglected, corresponding to the dark grey regions in Fig.~\ref{fig:RI}. In this case, the regime $f/v \lesssim 3.5$ is excluded for the Dirac scenario. 

\subsection{CMB reionization}
DM annihilation into ionizing particles in the post-CMB epoch may alter the residual ionization fraction, leading to detectable modifications of CMB polarization and temperature spectra. To a good approximation (see e.g. Ref.~\cite{Finkbeiner:2012sf}), these effects may be characterized by the redshift-independent quantity
\begin{equation}
p_{\textrm{ann}} =  g\frac{f_{\textrm{eff}}\langle \sigma v \rangle}{m_{\textrm{DM}}}\,,
\end{equation}
where $f_{\textrm{eff}} <1$ is the energy deposition efficiency of the ionizing final state particles and the prefactor $g$ encodes the differing degrees of freedom for Majorana or Dirac DM: $g = 1$ for Majorana and $1/2$ for Dirac. Current constraints from Planck `TT+TE,EE+lowP' polarization and temperature data provide a bound $p_{\textrm{ann}} < 4.1 \times 10^{-28}$\,cm$^3$s$^{-1}$GeV$^{-1}$  \cite{Ade:2015xua}. For photonic final states, one typically expects $f_{\textrm{eff}} \sim \mathcal{O}(1)$ \cite{Slatyer:2015kla,Slatyer:2015jla}. In Fig.~\ref{fig:RI} we show the corresponding exclusion regions for both Majorana and Dirac scenarios, for the maximal case that $f_{\textrm{eff}} = 1$. At present, the DM production contours are not excluded by this CMB reionization bound, although they may be probed by Planck in the cosmic variance limit.

\subsection{Direct detection}
\begin{figure}[ht]
	\centering\includegraphics[width=10cm]{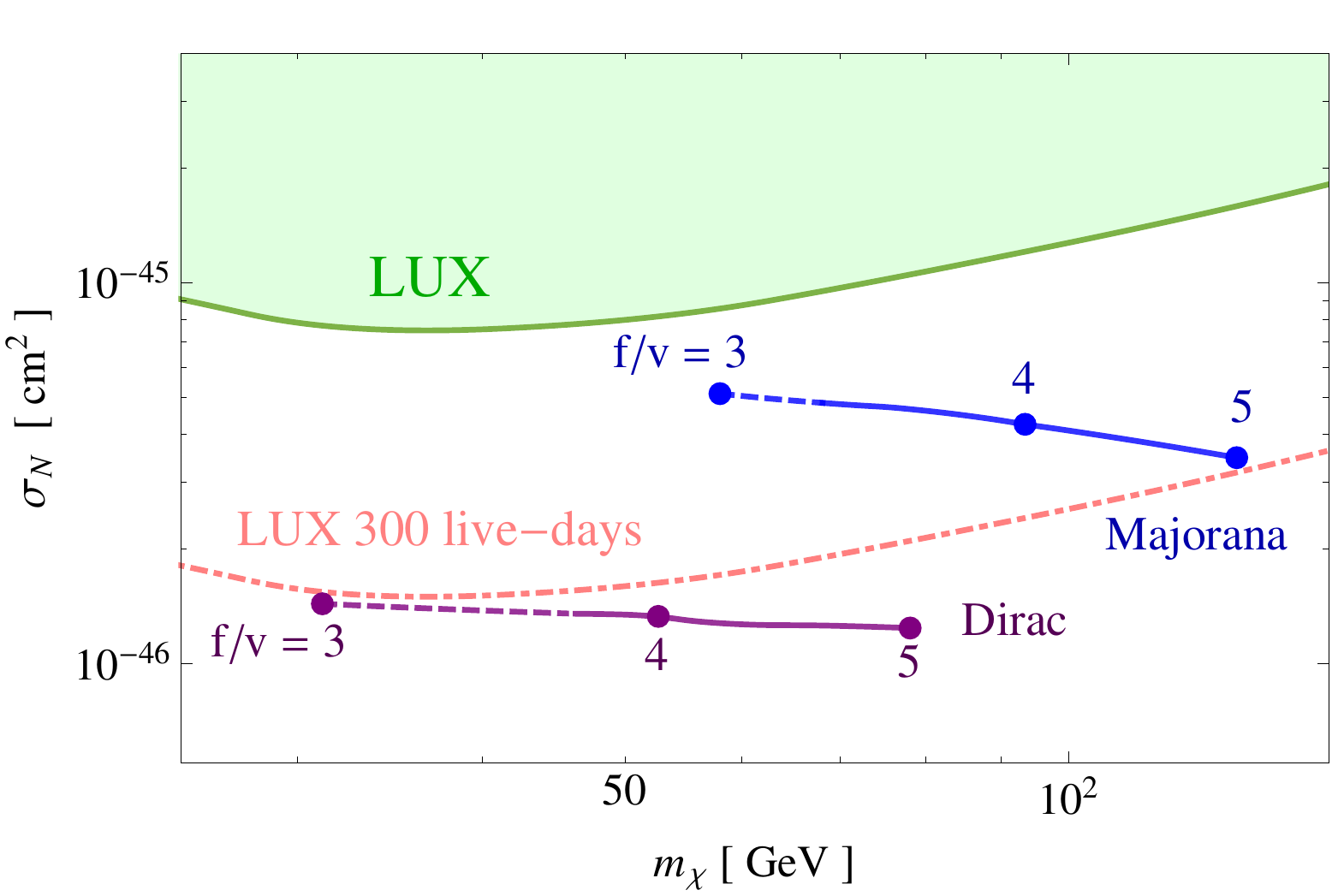}
	\caption{Nuclear scattering cross section for both the Dirac (purple) and Majorana (blue) DM scenarios, fixing $\Omega_{\psi,\,\chi} = \Omega_{\textrm{DM}}$ and $x_f = 20$. Integral values of $f/v$ are marked by filled circles. Dashed sections of the Majorana and Dirac contours correspond to the regions excluded by the Higgs invisible width, assuming $y_{b'}=y_b$. The current LUX bound (green) and the projected sensitivity for LUX at 300 live-days (red dot-dashed) are also included.}
	\label{fig:DD}
\end{figure}

DM particles in the twin sector can scatter with SM quarks through the Higgs portal, which may generate signals at indirect detection experiments. The scattering cross section to protons and neutrons is 
\begin{equation}
\sigma_{N}=\frac{1}{\pi}\bigg[\frac{m_{\textrm{DM}}\,v}{f^2}\bigg]^2g_{HN}^2\frac{\mu_{{\textrm{DM}} p}^2}{m_{h}^4}.
\end{equation}
We take the form factor $g_{HN}\simeq 1.2\times 10^{-3}$ for both proton and neutron from lattice studies \cite{Belanger:2008sj,Crivellin:2013ipa}. In Fig.~\ref{fig:DD} we plot the scattering cross section of both the Dirac and Majorana DM particles, with the mass and $f/v$ constrained to produce the full DM relic abundance, according to eqs.~\eqref{eqn:FVDM}, and assuming freeze-out temperature $x_f = 20$. The scattering cross-section decreases if $x_f$ is correspondingly reduced. The cross section of DM signals are well below the current LUX bound \cite{Akerib:2013tjd}. Anticipated future sensitivities \cite{Horn:2015ssa} may, however, have sufficient reach to probe this scattering.

\subsection{BBN bounds}
The twin and SM sectors typically decouple at the GeV temperature epoch, before confinement of the strong sectors. Without a twin baryon asymmetry, and without light twin leptons or radiation, at late times the entropy and energy density of the twin sector is deposited dominantly into the lightest hadron, here the $\twin{\pi}^0$. In order to avoid a $\twin{\pi}^0$ matter-dominated phase at or beyond the BBN epoch, or overclosure by $\twin{\pi}^\pm$, we therefore require the twin pion to decay into SM degrees of freedom before BBN (see similarly \cite{Farina:2015uea}), i.e.
\begin{equation}
\tau_{\twin{\pi}^0} < \tau_{\textrm{BBN}} \sim 1\,\mbox{s}\,.
\end{equation}
One may be concerned that the stable $\twin{\pi}^\pm$ may still freeze out from the hadronic plasma with a significant relic abundance, since the isospin-breaking induced mass splitting with the $\twin{\pi}^0$ is small. We therefore now verify that the $\twin{\pi}^\pm$ relic abundance is negligible. 

\begin{figure}[t]
	\centering\includegraphics[width=7cm]{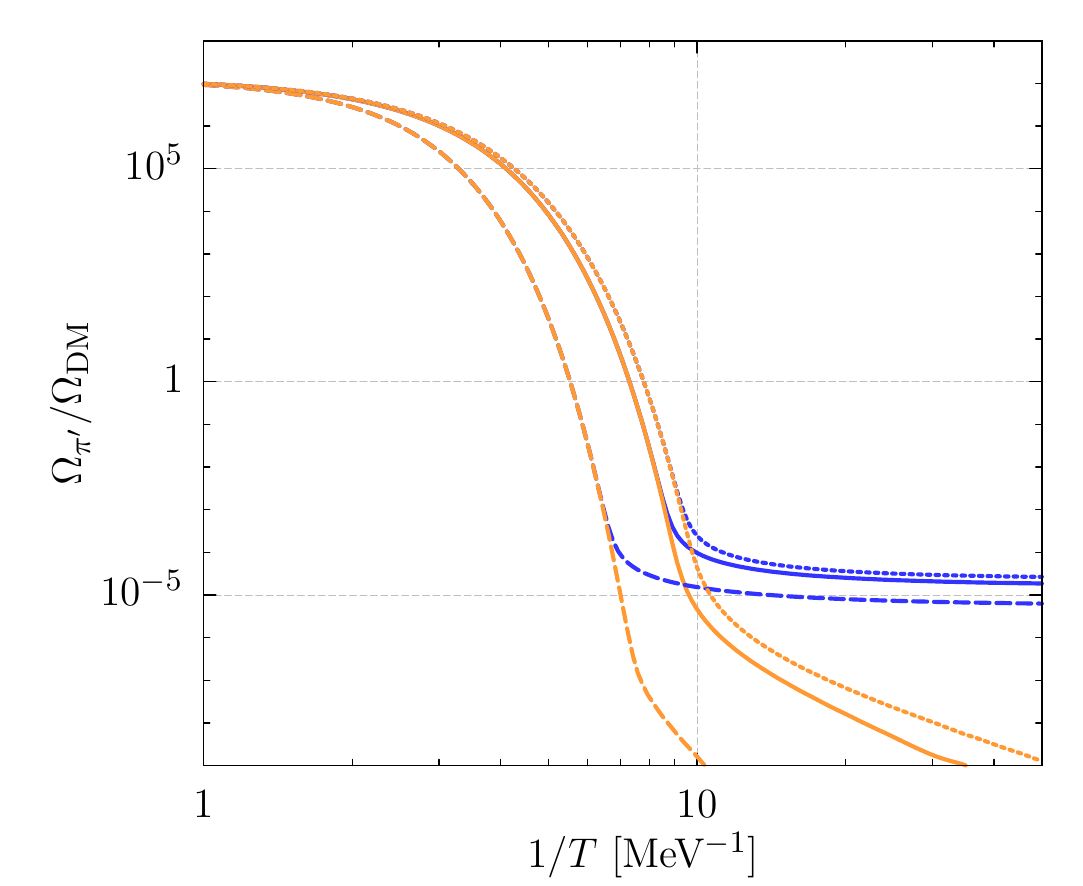}
	\caption{Relic abundances for $\twin{\pi}^0$ (gold) and $\twin{\pi}^\pm$ (blue), with mass splittings $\varepsilon = 10^{-5}$ (dotted), $10^{-4}$ (solid), and $10^{-3}$ (dashed). The twin pion mass $\bar{m}_{\twin{\pi}}=2\bar{m}_{\pi} $, the decay constant $f_{\twin{\pi}} = f_{\pi}$, and pion lifetime $\tau_{\twin{\pi}^0} = 1$\,s. }
	\label{fig:RA}
\end{figure}

If the twin sector remains in kinetic equilibrium with the SM sector, the pertinent Boltzmann equations for $\twin{\pi}^{\pm}$ freeze out are
\begin{align}
\frac{d Y_+}{d x} & = - \frac{x s \langle \sigma v \rangle_{+ \rightarrow 0}}{H(\bar{m}_{\twin{\pi}})} \Big[ Y_+^2(x) - Y_0^2(x) \Big]\,,\nonumber \\
\frac{d Y_0}{dx} & = -2\frac{x s \langle \sigma v \rangle_{0\rightarrow +}}{H(\bar{m}_{\twin{\pi}})} \Big[ Y_0^2(x) - Y_+^2(x) \Big] - \frac{x\Gamma_{\twin{\pi}^0}}{H(\bar{m}_{\twin{\pi}})} \Big[Y_0(x)  - Y^{\textrm{eq}}_0(x)\Big]\,. \label{eqn:BE}	
\end{align}
Here we assume the yield $Y_+ = Y_-$, and define the mean square pion mass $\bar{m}^2_{\twin{\pi}} \equiv (m_{\twin{\pi}^+}^2 + m_{\twin{\pi}^0}^2)/2$ and $x \equiv \bar{m}_{\twin{\pi}}/T$. The pion-pion scattering matrix element $ \simeq (s-m_{\twin{\pi}}^2)/f_\pi^2$ \cite{Weinberg:1996kr}, whence one may show the forward and inverse thermally averaged cross-sections
\begin{equation}
\langle \sigma v \rangle_{+\rightleftarrows 0} \simeq \frac{9}{128\pi^2} \frac{\varepsilon x \bar{m}^2_{\twin{\pi}}}{ f_{\twin{\pi}}^4} K_1(2x)K_1(\varepsilon x)e^{x(2\pm \varepsilon)}\,,
\end{equation}
in which $\varepsilon \equiv  (m_{\twin{\pi}^+}^2 - m_{\twin{\pi}^0}^2)/ (m_{\twin{\pi}^+}^2 + m_{\twin{\pi}^0}^2) > 0$.  Solving eqs.~\eqref{eqn:BE}, in Fig.~\ref{fig:RA} we show the thermal history of the charged twin pion relic abundances for a range of mass splittings $\varepsilon$. Even in the $\varepsilon \ll 1$ regime, the relic abundance for the $\twin{\pi}^\pm$ is $\Omega_{\pi'^{\pm}}\sim 10^{-5}\;\Omega_{\textrm{DM}}$, so there is no significant twin pion thermal relic. During the recombination epoch, charged twin pions with this small of an abundance annihilate into SM photons with thermal cross section $\langle\sigma v\rangle\sim10^{-32}$ cm$^3$s$^{-1}$. This is well below the current upper bound $\sim10^{-29}$cm$^{-3}$s$^{-1}$ for $m_{\pi'}\simeq 100$\,MeV, arising from measurements of cosmic microwave background (CMB) anisotropies \cite{Slatyer:2015jla,Slatyer:2015kla}.

Note moreover that for a twin pion of mass $m_{\twin{\pi}} \sim 100$~MeV, number changing processes such as $4\pi'\to2\pi'$ remain efficient until $m_{\twin{\pi}}/T \sim 10$ (see e.g. \cite{Hochberg:2014dra}). If the twin sector is kinetically decoupled from the SM, then such interactions may cause the twin sector temperature to rise exponentially compared to the SM sector. This keeps the $\twin{\pi}^\pm$ in equilibrium with $\twin{\pi}^0$ until later times, further suppressing the charged pion relic abundance.

\section{Twin pion decay}
\label{sec:TPDM}

In the absence of twin hypercharge interactions, the twin pion can nominally only decay by mixing with a $0^{++}$ isospin singlet state, which can subsequently decay through the Higgs portal to SM degrees of freedom. Such mixing is however heavily suppressed by both parity and the small twin isospin violating Yuwaka coupling, $y_{\twin{u}}-y_{\twin{d}}$. For instance, the twin-isospin and parity-violating $\twin{\pi}^0\twin{G}_{\mu\nu} \twin{G}^{\mu\nu}  /f_{\twin{\pi}}$ operator enters only at least at two loop order \cite{Czarnecki:1996rx}, since the only source of parity violation is the twin electroweak interactions, assuming a negligible $\twin{\theta_{\textrm{qcd}}}$. The twin pion decay amplitude is thus heavily suppressed, leading to lifetimes well in excess of $\tau_{\textrm{BBN}}$. In order to allow the $\twin{\pi}^0$ to decay before BBN, it is therefore necessary to introduce an additional portal between the twin and SM sectors, 

In this section, we first perform an effective operator analysis for such interactions. For $m_{\pi'^0}<3 m_{\pi^0}$, this analysis reveals the need for a UV completion of the additional twin-SM portal below the TeV scale. We subsequently present a sample UV completion and analyze the corresponding constraints on its parameter space, both from direct searches at the LHC as well as from precision flavor measurements. Alternatively, in the regime that $3 m_{\pi^0}<m_{\pi'^0}<2m_{K}$, hadronic decays of the twin pion to three pion final states -- i.e. $\twin{\pi}^0 \to 3\pi$ --  predominantly produce soft photons, and are sufficiently fast that no extra twin-SM mediator is required below the twin Higgs cutoff, $\Lambda \sim 5$~TeV. Hence in this regime the hadrosymmetric twin Higgs framework does not require any near-term detectable phenomenology in order for the twin pions to decay before BBN. In the regime $m_{\twin{\pi}^0} > 2m_K$, the available SM hadronic decay channels of the twin pion become large in number. This case would also require a rather large $f/v\gtrsim 10$ from eq.~\eqref{eqn:TPM}, and we therefore do not consider it in this work.

\subsection{Effective field theory analysis}
\label{sec:decayrates}
The twin and SM confinement scales are readily probed at current collider experiments, so we therefore must consider a (twin) quark effective theory for the twin pion decay. The twin pion decay must be mediated by a twin isospin-violating current, to avoid suppression of the decay rate by the small $y_{\twin{u}}-y_{\twin{d}}$ coupling. Under the approximate $\mathbb{Z}_2$ symmetry between the SM and twin quark sectors, this current should similarly couple to SM quarks rather than leptons, and be SM isospin-violating, too. In this framework, then, the $\twin{\pi}^0$ decay is maximal in the case that it mixes with the lightest SM pseudoscalar in a non-trivial isospin representation, that is, the $\pi^0$.

Gauge kinetic mixing portals with SM hypercharge cannot induce $\twin{\pi}^0$ tree-level diphoton or dilepton decays, because they do not mix longitudinal modes of heavy vector bosons, and therefore generate an insufficiently fast twin pion decay. The lowest dimensional viable portals are then dimension-six current-current interactions.  We consider vectorial or axial neutral currents, and further require them to be flavor-conserving to avoid precision flavor constraints. Portals involving the $V-A$ current $\twin{Q}^\dagger \sigma_\mu \twin{Q}$ do not generate the required isospin breaking in the twin sector. We therefore focus only on $V+A$ interactions, which couple to both SM and twin right-handed quarks. 

Defining the current 
\begin{equation}
J^{q\,\mu}_{V+A} = q^\dagger \sigma^\mu q\,
\end{equation}
for right-handed $q = u^i$,$d^i$,$\twin{u}^i$ or $\twin{d}^i$, pion-twin-pion mixing is generated by the effective operator
\begin{equation}
\label{eqn:EFTQ}
\frac{1}{\Lambda_P^2} \big(J^{\twin{u}}_{V+A} - J^{\twin{d}}_{V+A} \big)\cdot \big(J^{u}_{V+A} - J^{d}_{V+A} \big)\,,
\end{equation}
where $\Lambda_P$ is the scale of the twin-SM portal. This operator is parity violating but CP conserving, and therefore permits a $J^{PC} =  0^{-+}$ state to mix with either $0^{-+}$ or $0^{+-}$ states.

Before proceeding, it should be mentioned that there are several alternative portals, other than a current-current operator.  First, one might generate the $\twin{\pi}^0$ decay with a spin-0 mediator coupling to $\twin{Q}\twin{u}^c$. Such a mediator must belong to a twin electroweak doublet. This could be implemented by extending both the Higgs and twin Higgs sectors, for example in a twin two Higgs doublet model. However, since this additional Higgs doublet must couple to the quark sector, the Glashow--Weinberg condition is violated and one generically expects large deviations in precision flavor experiments. While this may be interesting direction for further work, we do not pursue this option further here. Second, one could instead consider explicitly breaking the SM-twin $\mathbb{Z}_2$ exchange symmetry, by allowing a $\twin{\pi}^0$ decay channel to leptons, mediated by the effective operator 
\begin{equation}
\label{eqn:EFTL}
\frac{1}{\Lambda_P^2} \big(J^{\twin{u}}_{V+A} - J^{\twin{d}}_{V+A} \big)^\mu\; \ell^\dagger_R \sigma_\mu \ell_R\,,
\end{equation}
where $\ell = \mu$ or $e$ for the $\twin{\pi}^0$ masses under consideration herein. 

\subsection{Decay rate and branching ratio estimates}
\label{sec:DRBRE}
The amplitude for $\twin{\pi}^0 \to \textrm{SM}$ decay, as generated by the operator~\eqref{eqn:EFTQ}, receives its dominant contribution from the off-shell $\pi^0$ channel, i.e $\twin{\pi}^0 \to \pi^{0*} \to \textrm{SM}$. The next lightest isospin triplet pseudoscalar is the much heavier $\pi(1300)$, and we neglect Yukawa-suppressed SM isospin violating effects. To estimate the decay rates, observe that the mixing amplitude
\begin{equation}
\big\langle \pi^{0*}(p) \big| \big(J^{u}_{V+A} - J^{d}_{V+A} \big)\cdot\big(J^{\twin{u}}_{V+A} - J^{\twin{d}}_{V+A}\big)\big| \twin{\pi}^0(\twin{p})\big\rangle \simeq f_{\twin{\pi}}f_\pi p\cdot\twin{p}\,,
\end{equation}
so that
\begin{align}
\Gamma[\twin{\pi}^0 \to \textrm{SM} ] 
& \simeq \frac{m^4_{\twin{\pi}^0}f^2_{\twin{\pi}}f^2_\pi}{\Lambda_P^4} \frac{\Gamma[\pi^{0*} \to \textrm{SM} ]}{(m^2_{\twin{\pi}^0} - m^2_{\pi^0})^2 + m_{\pi^0}^2\Gamma_{\pi^0}^2}\notag\\
& \simeq \lambda^4 (f/v)^2 \frac{f^4_\pi}{\Lambda_P^4}\frac{\Gamma[\pi^{0*} \to \textrm{SM} ]}{(\lambda f/v -1)^2 + \Gamma_{\pi^0}^2/m_{\pi^0}^2} \,,\label{eqn:TPDR}
\end{align}
in which we have applied the scalings in eqs.~\eqref{eqn:THM} and \eqref{eqn:TPM}. Thus the decay rate and branching ratios for the $\twin{\pi}^0$ are immediately informed by estimating the corresponding partial widths of an off-shell $\pi^{0*}$, with mass $m_{\twin{\pi}^0}$. 

These partial widths are, as usual, subject to various kinematic thresholds as $m_{\twin{\pi}^0}$ is varied. For instance, if $m_{\twin{\pi}^0} <2m_{\pi^0}$, the twin pion decays predominantly to $\gamma\gamma$ through the chiral anomaly. For $2m_{\pi^0} < m_{\twin{\pi}^0} < 3m_{\pi^0}$, the $\pi^+\pi^-\gamma$ final state becomes available. The kinematically accessible $\pi'^0\rightarrow\pi^+\pi^-$ or $2\pi^0$ mode is, however, both parity and CP violating. It is therefore not mediated by the CP-conserving operator \eqref{eqn:EFTQ}, and is negligible. Finally, for $m_{\twin{\pi}^0}>3m_{\pi^0}$, the twin pion can decay to three SM pions and further exclusive hadronic modes. These strong decays are not suppressed by isospin violation, and are therefore expected to be much faster than the electromagnetic decay modes.

To estimate the branching ratios of the twin pion with eq.~\eqref{eqn:TPDR}, we adapt data from the SM $\pi^0$ and $\eta$ -- the neutral pseudoscalars closest in mass to the $m_{\twin{\pi}^0}$ regime of interest --  rescaling the partial widths according to their explicit mass dependence. To adapt the $\eta$ data, we take care to also rescale by Clebsch-Gordan coefficients associated to decays of an isospin triplet rather than a singlet, as well as including isospin violating effects and $\eta$-$\eta'$ mixing. (We emphasize that the $\eta'$ here is the SM hadron, not a twin state.) One finds, in a particular basis of isospin invariants, that
\begin{align}
\frac{\Gamma[ \pi^{0*} \to \pi^+\pi^-\gamma] }{ \Gamma[ \pi^{0*} \to 2\gamma]}  & = \bigg\{\frac{\sqrt{6} - 6r_2 \tan\phi}{\sqrt{6} - 2r_1 \tan\phi}\bigg\}^2\;\frac{\Gamma[ \eta^* \to \pi^+\pi^-\gamma] }{ \Gamma[ \eta^* \to 2\gamma]}\,,\notag\\[5pt]
\frac{\Gamma[ \pi^{0*} \to 3\pi] }{ \Gamma[ \pi^{0*} \to 2\gamma]}  & = \frac{1}{\delta_{\textrm{SM}}^2}\bigg\{ \frac{4 - 4\sqrt{6}r_2 \tan\phi}{6 r_3 - 
	\sqrt{6} r_3 \tan\phi}\bigg\}^2\;\frac{\Gamma[ \eta^* \to 3\pi] }{ \Gamma[ \eta^* \to 2\gamma]}\,,
\end{align}
where $\phi$ is the $\eta$-$\eta'$ mixing angle, $\delta_{\textrm{SM}} \sim 1\%$ is the SM isospin breaking scale, and $r_{1,2,3}$ are unknown ratios of hadronic matrix elements. Estimates provide $\phi \simeq 40^\circ$ (see e.g. \cite{Ambrosino:2009sc}). (For the ideal mixing case that the $\eta'$ is a pure $s\bar{s}$ state, $\phi = \tan^{-1}\sqrt{2} \simeq 55^\circ$.) Taking a na\"\i ve average over $\mathcal{O}(1)$ values for $r_{1,2,3}$, for $ 2 m_{\pi} < m_{\twin{\pi}^0} < 3 m_{\pi}$ one estimates
\begin{equation}
\label{eqn:PPPG}
\frac{\Gamma[ \pi^{0*} \to \pi^+\pi^-\gamma] }{ \Gamma[ \pi^{0*} \to 2\gamma]} \sim 6. \bigg[\frac{m_{\twin{\pi}^0}}{m_\eta}\bigg]^4\;\frac{\Gamma[ \eta \to \pi^+\pi^-\gamma] }{ \Gamma[ \eta \to 2\gamma]}\,.
\end{equation}
In eq.~\eqref{eqn:PPPG} we assume that the $\eta \to \pi^+\pi^-\gamma$ process is vector-meson dominated, so that it can be thought of as $\eta \to (\rho^{0*}\to \pi^+\pi^-)\gamma$, and hence its rate na\"\i vely scales as $m_\eta^7$. We do not include the phase space effects that smoothly turn off the $\twin{\pi}^0 \to \pi^+\pi^- \gamma$ rate nearby to the $m_{\twin{\pi}^0} = 2m_\pi$ threshold. For $m_{\twin{\pi}^0} > 3m_{\pi}$, similarly one estimates
\begin{equation}
\label{eqn:P3P}
\frac{\Gamma[ \pi^{0*} \to 3\pi] }{ \Gamma[ \pi^{0*} \to 2\gamma]} \sim 3.\times 10^4\bigg[\frac{m_{\twin{\pi}^0}}{m_\eta}\bigg]^2 \;\frac{\Gamma[ \eta \to 3\pi] }{ \Gamma[ \eta \to 2\gamma]}\,,
\end{equation}
where the $3\pi$ state is either $\pi^+\pi^-\pi^0$ or $3\pi^0$. The $\eta \to 3\pi$ decay na\"\i vely scales as $m_\eta^5/f_{\eta}^4$, whence the mass scaling dependence in eq.~\eqref{eqn:P3P}. Note that relative Clebsch-Gordan coefficients, combinatorics and phase space symmetry factors imply that $\Gamma[ \pi^{0*}/\eta^* \to 3\pi^0]/\Gamma[ \pi^{0*}/\eta^* \to \pi^+\pi^-\pi^0] \simeq 3/2$. Thus in the $\twin{\pi}^0 \to 3\pi$ decay mode, one expects the photon versus lepton production ratio to be approximately $11:4$. Therefore, in the regime $3m_{\pi^0} < m_{\twin{\pi}^0} < 2m_{K}$ one expects twin pion decays to predominantly produce a high multiplicity of comparatively soft photons.

\begin{figure}[t]
	\includegraphics[width=0.48\textwidth]{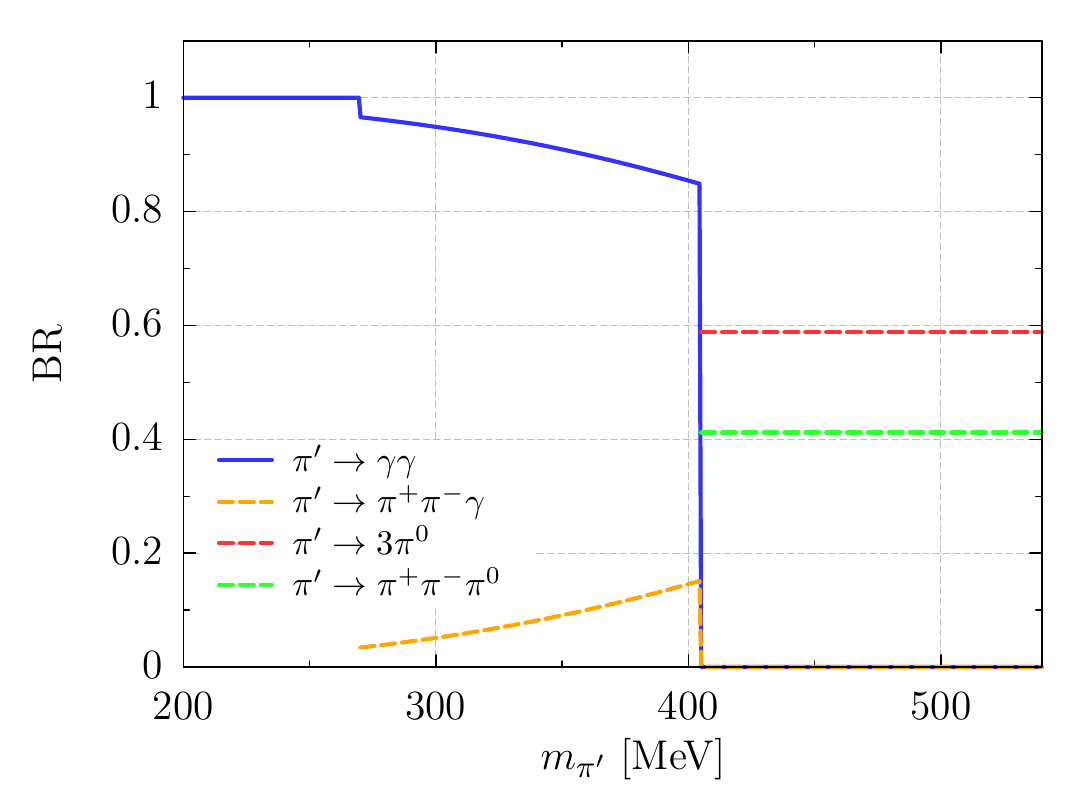}\hfill
	\includegraphics[width=0.48\textwidth]{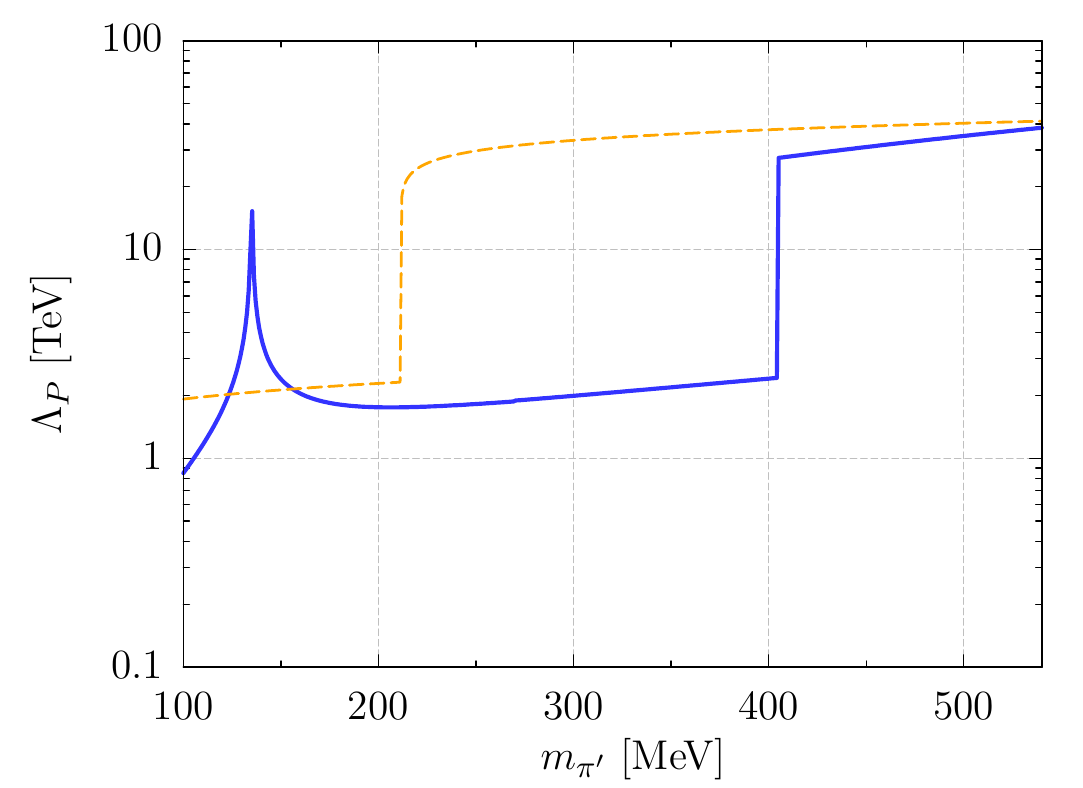}
	\caption{Left: Schematic branching ratios as a function of the twin pion mass for decays mediated by the operator \eqref{eqn:EFTQ}. Phase space effects near thresholds are omitted. Right: Maximum mediator scale $\Lambda_P$ needed to satisfy BBN bound on twin pion lifetime ($\tau<1$ s), as a function of $m_{\pi'}$, for decay through the hadronic (blue) and leptonic (orange dashed) portals \eqref{eqn:EFTQ} and \eqref{eqn:EFTL} respectively.}
	\label{fig:decaymodes}
\end{figure}

The corresponding estimated twin pion branching ratios as a function of $m_{\twin{\pi}^0}$ are shown in the left-hand panel of Fig.~\ref{fig:decaymodes}, for $f_{\twin{\pi}}=f_\pi$. We see there that for $m_{\twin{\pi}^0} < 2m_{\pi}$, the diphoton rate dominates as expected. For $2m_{\pi} < m_{\twin{\pi}^0} < 3m_{\pi}$ the diphoton mode continues to dominate the $\pi^+\pi^-\gamma$ decay mode, while for $m_{\twin{\pi}^0} > 3m_{\pi}$, the purely hadronic $3\pi$ modes dominate overwhelmingly. In the righthand panel of Fig.~\ref{fig:decaymodes} we show the maximum effective scale $\Lambda_P$ under which the twin pion decays before BBN -- i.e. $\tau_{\twin{\pi}^0} \lesssim 1$\,s -- in these different kinematic regimes. We see in Fig.~\ref{fig:decaymodes} that to satisfy the BBN bound for $m_{\twin{\pi}^0} < 3m_{\pi}$, the portal \eqref{eqn:EFTQ} generically requires a UV completion nearby the TeV scale, while for $m_{\twin{\pi}^0} > 3m_\pi$, the BBN bound is satisfied generically for $\Lambda_P$ well above the scale of twin Higgs effective theory, $\Lambda \sim 5$ to $10$~TeV. Since the LHC probes TeV energies, an effective operator approach is therefore insufficient to study possible collider constraints for $m_{\twin{\pi}^0} < 3m_{\pi}$. We shall therefore present a sample UV completion for this regime in the next section, as well as its corresponding experimental constraints.  

Finally, if the leptonic portal \eqref{eqn:EFTL} is available, then the twin pion may also decay rapidly to leptons, subject to the usual chiral suppression. In particular, 
\begin{equation}
\big\langle  \ell^+\ell^- \big|\big(\ell^\dagger_R \sigma_\mu \ell_R\big)\big(J^{\twin{u}}_{V+A} - J^{\twin{d}}_{V+A}\big)^\mu \big| \twin{\pi}^0(p)\big\rangle \simeq f_{\twin{\pi}} \,\bar{u}(p_{\ell^-}) \slashed{p}' P_R v(p_{\ell^+})\,,
\end{equation}
whence one may compute the decay rate directly, viz.
\begin{equation}
\Gamma[\twin{\pi}^0 \to \ell^+ \ell^- ] \simeq \frac{ f_{\twin{\pi}}^2 m_{\ell}^2 m_{\twin{\pi}^0}}{8 \pi \Lambda_P^4}\sqrt{1 - \frac{4m^2_\ell}{m_{\twin{\pi}^0}^2}}\,.
\end{equation}
Hence if kinematically accessible, the muon decay channel dominates, followed by a much weaker decay to electrons. In the righthand panel of Fig.~\ref{fig:decaymodes} we show also the maximum effective scale $\Lambda_P$ required to ensure $\tau_{\twin{\pi}^0} \lesssim 1$\,s via the leptonic portal \eqref{eqn:EFTL}. The muon decay channel is fast enough by itself that the twin pion lifetime BBN bound can be satisfied with $\Lambda_P \gtrsim 10$~TeV. Therefore, in this scenario no UV completion is required below the twin Higgs cutoff. (Alternatively, one could also modify the model to include gauged twin hypercharge, such that the twin pion can decay to two twin photons \cite{Farina:2015uea}. The twin photons then subsequently decay into SM leptons through the kinetic mixing portal.) However, this leptonic decay channel may introduce possible tensions with astrophysical bounds on high multiplicity muon production by DM annihilations \cite{Bergstrom:2013jra,Bringmann:2014lpa}. A full analysis of these astrophysical bounds is beyond the scope of this work, so we shall therefore not consider this channel further.

\subsection{A sample UV completion}
\label{sec:Zprime}
\enlargethispage{-2\baselineskip} 
A straightforward UV completion of the operator \eqref{eqn:EFTQ} can be achieved by charging the right-handed quarks under an additional broken $U(1)_X$ gauge interaction, such that $u$ ($\twin{u}$) and $d$ ($\twin{d}$) have opposite charges.\footnote{Here we choose to adhere to the spirit of neutral naturalness by not introducing additional, TeV-scale colored states. There exist, however, UV completions of the operator \eqref{eqn:EFTQ} if we lift this requirement. For example, a $t$-channel scalar interaction of the form $\xi^u_{ij}  \phi_u  Q_i'  \bar u_j+\xi^d_{ij} \phi_d  Q_i'  \bar d_j+\xi^u_{ij}   \phi_u'  Q_i   \bar u'_j+\xi^d_{ij}  \phi'_d  Q_i \bar d'_j$, in which the Yukawas $\xi$ may be dynamically aligned with the SM and twin Yukawas (see e.g. \cite{Knapen:2015hia}). The scalars $\phi$ must carry both twin and SM colors and therefore this UV completion suffers strong constraints from collider searches. Options of this type were also explored in a different, emerging-jet context in \cite{Schwaller:2015gea}. 
} 
That is, an interaction of the form
\begin{equation}
\label{eq:Zpmodel}
\mathcal{L} = g_X X_\mu \big( u^\dagger \sigma^\mu  u + \twin{u}^\dagger \sigma^\mu  \twin{u}\big) - g_X X_\mu \big( d^\dagger \sigma^\mu  d + \twin{d}^\dagger \sigma^\mu  \twin{d}\big)\,,
\end{equation}
where $X$ is the massive $U(1)_X$ gauge boson, and $g_X$ is the gauge coupling. Note also that charging the right-handed quarks in this manner introduces mixed $U(1)_X$-hypercharge instanton anomalies. However these can be cancelled with a small set of TeV-scale hypercharged anomalons. 

To accommodate the BBN bound on the twin pion lifetime for $m_{\twin{\pi}^0} < 3m_{\pi}$, from Fig.~\ref{fig:decaymodes} one requires
\begin{equation}
\label{eqn:BBNXM}
m_{X}/g_X = \Lambda_P \lesssim 2~\mathrm{TeV}\,.
\end{equation} 
We consider the range $10\; \mathrm{MeV}<m_{X}<2\;\mathrm{TeV}$ and allow $g_X$ to vary to accommodate this BBN constraint.

The $U(1)_X$ charge assignments imply that the twin and SM Yukawas must be modified to the form
\begin{equation}
\label{UVcompschannel}
\tilde y_u \frac{\phi}{\Lambda^u_\phi} H \bar{Q} u +\tilde y_d \frac{\phi^\dagger}{\Lambda^d_\phi} H^\dagger \bar{Q}  d+\twin{\tilde{y}}_u \frac{\phi}{\Lambda^u_\phi} \twin{H} \twin{\bar{Q}}  \twin{u} +\twin{\tilde{y}}_d \frac{\phi^\dagger}{\Lambda^d_\phi} \twin{H}^\dagger \twin{\bar{Q}}  \twin{d}\,,
\end{equation}
where $\phi$ is a scalar charged under the $U(1)_X$, and the interaction scales $\Lambda^{u,d}_\phi$ may be different for up and down-type couplings, but are at least as high as the twin Higgs cut-off, $\Lambda$. The SM Yukawas are then generated by $y_{u,d}=\tilde y_{u,d} \langle \phi \rangle /\Lambda^{u,d}_\phi$. The vev $\langle \phi \rangle$ contributes to the mass of the $X$, but may not be the only contribution in a complete model, and we therefore keep the mass of the $X$, $m_{X}$, as an independent parameter.

If the $X$ couples universally to all three quark generations, the requirement of an $\mathcal{O}(1)$ top Yukawa implies that $\langle \phi \rangle \sim \Lambda^u_\phi$. Assuming $\Lambda^{u,d}_\phi\gtrsim \Lambda \sim 5$~TeV, there will be a region of parameter space with relatively large $g_X$ and low $m_{X}$ where some additional fine-tuning is needed to ensure that the $X$ remains light compared to $g_X\langle \phi \rangle$. But, as we will show in the following section, the tuned region of parameter space for $m_{X} >1$~GeV happens to be outside the experimentally allowed regions, and therefore this tuning is not of great concern.

An alternative is to imagine that the $U(1)_X$ does not couple to the third quark generation, but rather couples in a horizontal fashion to the first two generations only. In this case it is possible to have $\langle \phi \rangle \ll \Lambda^{u,d}_\phi$ -- one only requires $\langle \phi \rangle/\Lambda^{u,d}_\phi \gtrsim \sqrt{2}m_{c,s}/v$ -- and the $X$ may naturally be light in the full parameter space. The horizontal nature of the coupling of $\phi$ to quarks, however, introduces extra flavor changing neutral currents through the $\phi$ order parameter, as well as extra CP violating phases.\footnote{With a $2+1$ horizontal coupling, the $Q$, $u^c$ and $d^c$ kinetic terms have a $U(3)\times U(2)^2 \times U(1)^2$ flavor symmetry which is broken to $U(1)_B$ by the Yukawas. This corresponds to $13$ ($5$) broken imaginary (real) generators, and therefore introduces four more physical mixing angles and four more physical phases compared to the SM.} Requiring that $\langle \phi\rangle /\Lambda^{u,d}_\phi \gtrsim m_{c,s}/v$ and assuming $M_\phi \sim \langle \phi \rangle \sim \Lambda_P$, one then deduces from eq.~\eqref{eqn:BBNXM} that the scale of the new flavor-violating interactions $\Lambda^u_\phi \lesssim \Lambda_P^2/m_c \lesssim 3 \times 10^3$~TeV and $\Lambda^d_\phi \lesssim \Lambda_P^2/m_s \lesssim 4 \times 10^4$~TeV, for up and down-type flavor violating neutral processes respectively. These estimates are in $\mathcal{O}(10)$ tension with existing bounds \cite{Isidori:2010kg}, but can perhaps be ameliorated by embedding the model in a more complete theory of flavor. Since the twin Higgs setup itself requires a UV completion near 5 TeV already, we shall not attempt to construct such a model. Instead we shall keep these flavor challenges in mind, and employ the horizontal model hereafter merely to illustrate how the various collider constraints depend on the properties of the UV completion of the portal \eqref{eqn:EFTQ}.

One might have also considered charging the Higgs under the $U(1)_X$, thereby eliminating the need for the spurion $\phi$ in the quark sector. (Since the leptons are uncharged under $U(1)_X$, a spurion would still be needed for the lepton Yukawas.) This approach is, however, plagued with problems, as the Higgs vev now induces a tree-level mixing between the $X$ and both the SM $Z$ and twin $\twin{Z}$ bosons. This induces strong tensions with both electroweak precision tests and DM direct detection experiments. In particular for the case where the dark matter is a Dirac fermion, $\psi$, the mixing induces the four-fermi operator $\psi^\dagger\gamma_{\mu}\psi\,q^\dagger\gamma^{\mu}q/\Lambda^2_{P}$. This is ruled out by current direct detection experiments, given the BBN bound \eqref{eqn:BBNXM}. We therefore do not consider this approach in this work.

\subsection{Model constraints}
\label{sec:MC}

At high $m_{X}$ -- above several hundred GeV -- the only relevant collider constraints come from LHC searches, while for low and intermediate $m_{X}$ -- below and above 1~GeV respectively -- a variety of experiments at the intensity frontier play a role as well. In what follows, we separately consider the mass ranges $m_{X}>300$~GeV,   $300~\mbox{GeV}>m_{X}>1$~GeV and  $m_{X} <1$~GeV, at the following benchmark points:
\begin{equation}
\label{eqn:MFB}
\begin{array}{lll}
\text{benchmark 1:}\quad\quad&m_{\pi'^0}=250\text{ MeV,}\quad \quad f_{\pi'}=f_\pi\,,\\
\text{benchmark 2:}\quad\quad&m_{\pi'^0}=350\text{ MeV,}\quad \quad f_{\pi'}=2f_\pi\,,
\end{array}
\end{equation}
where the latter benchmark anticipates the best fit value for the galactic center $\gamma$-ray excess, discussed in Sec.~\ref{sec:indirect}.

$\bm{m_{X} > 300~\mbox{GeV} }$: Rescaling the dijet and monojet LHC bounds \cite{Chala:2015ama}, one finds that $m_{X} > 500$~GeV is presently already directly excluded. For lower masses and lower couplings these searches lose sensitivity, but the combination of the monojet bound with the BBN constraint on the twin pion lifetime still excludes $m_{X} \gtrsim 300$~GeV. 

\begin{figure}[t]
	\includegraphics[width=0.49\textwidth]{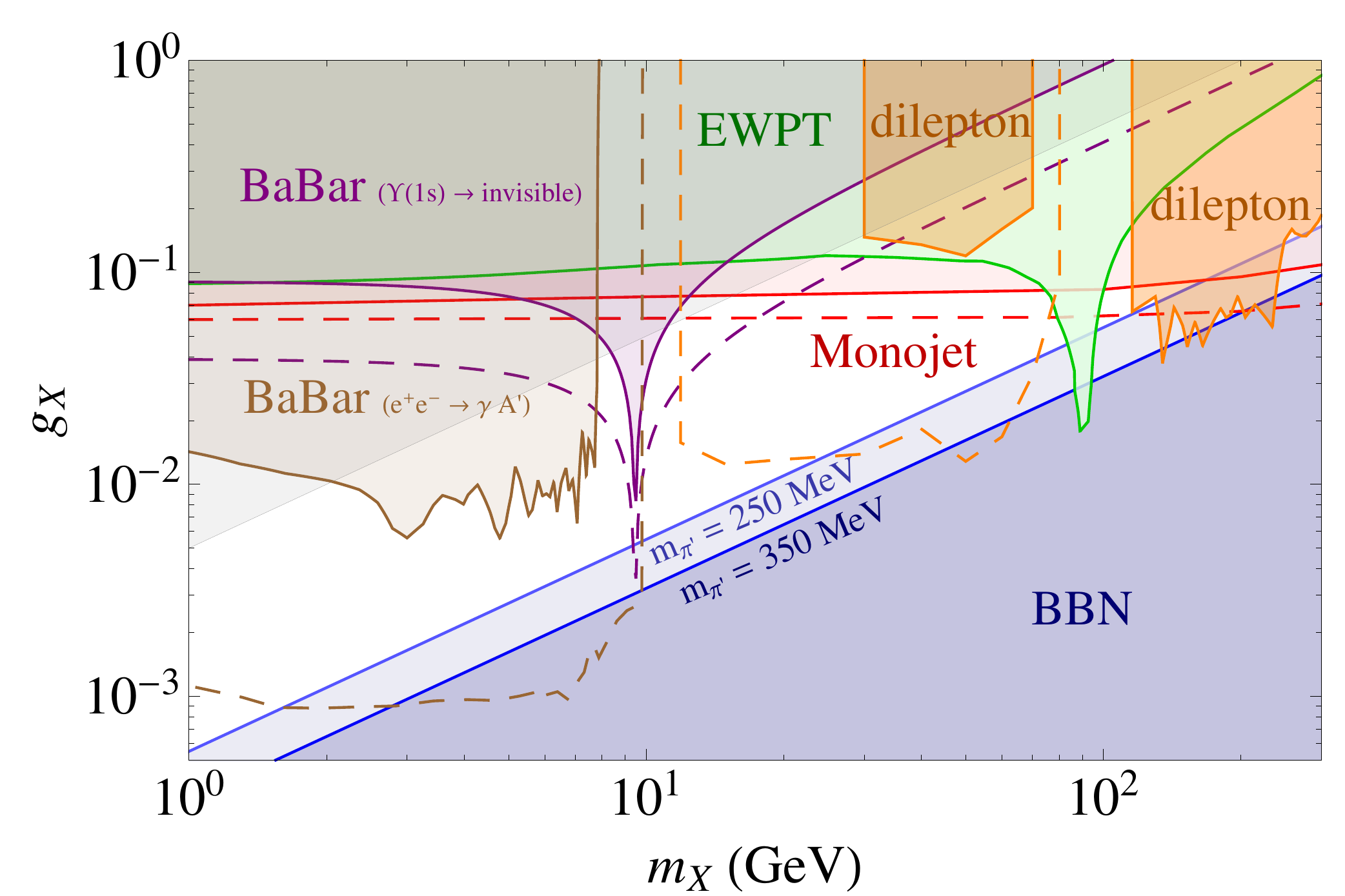}\hfill
	\includegraphics[width=0.49\textwidth]{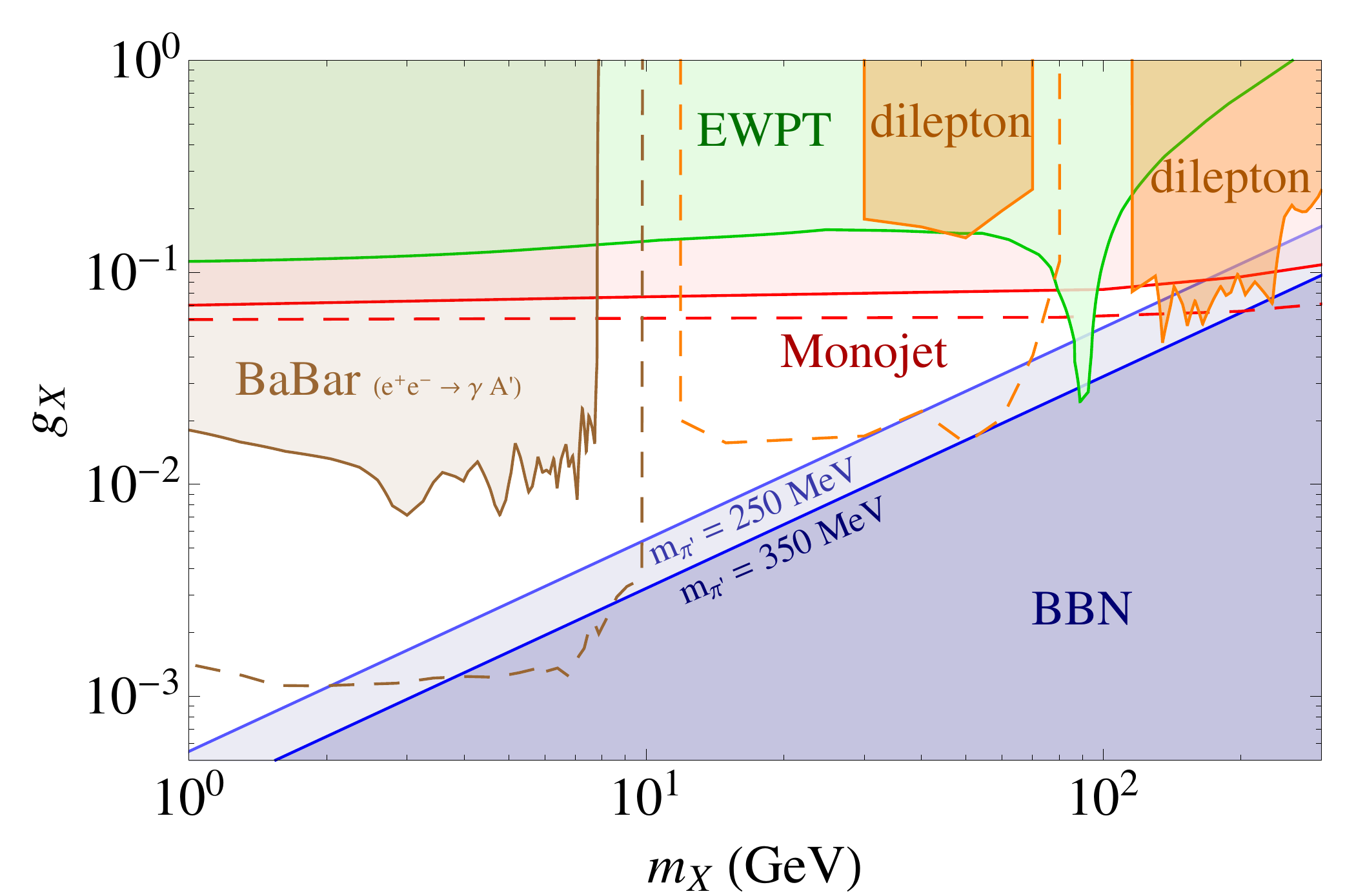}
	\caption{Constraints on the $m_{X}$--$g_X$ parameter space for a universal (left) and horizontal (right) $X$ coupling. See text for details. The light (dark) blue shaded  region displays the BBN bound for benchmark 1 (benchmark 2). Dashed lines correspond to projections of future experimental reach.  In the gray shaded region for the universal case, at least 10$\%$ fine-tuning is required to keep the $X$ light. }
	\label{fig:Zprimeconstraints}
\end{figure}

$\bm{1~\mbox{GeV} < m_{X} < 300~\mbox{GeV}}$: 
In this regime LHC monojet searches as well as a variety of dark photon searches at $B$ factories apply, but the BBN bound is the only bound which depends on the choice of $\{m_{\pi'^0},f_{\pi'}\}$. In Fig.~\ref{fig:Zprimeconstraints} we show all relevant constraints on the $\{m_{X}, g_X\}$ parameter space, together with the BBN bound \eqref{eqn:BBNXM}. Projections of future experimental reach are also shown. The most robust constraint is provided by a recasting the CMS monojet analysis \cite{CMS-PAS-EXO-12-048}, as indicated by the red line.  For this purpose we generated $pp\to j+X$ events using MadGraph 5$+$Pythia 6.4$+$PGS 2.4.3 \cite{Alwall:2014hca,Sjostrand:2006za,pgs} and FeynRules 2.0 \cite{Alloul:2013bka}, where we take the branching ratio of $X$ to twin quarks to be 50\%. The most important cut for this search on the missing transverse momentum $\slashed E_T>350$ GeV (see Ref.~\cite{CMS-PAS-EXO-12-048} for more details).  With the same background analysis as in Ref.~\cite{Primulando:2015lfa}, we obtain a $95\%$ CL upper bound on $g_X$, which has only a very mild dependence on $m_{X}$. For the $14$ TeV projection, shown by the red-dashed curve, we keep all the cuts in the $8$ TeV search except $\slashed E_T>550$ GeV. From the projection of a data-driven analysis \cite{Primulando:2015lfa}, we obtain this bound with $300$ fb$^{-1}$ of data and an estimated $3.4\%$ systematic uncertainty. Unless there are significant improvements in the reduction of uncertainties for the signal acceptance and selection efficiency, the systematic uncertainty on the $14$ TeV search is not expected to be much lower than the CMS study. Hence the improvement of this bound is very limited compared to the $8$ TeV search. The dijet constraints in this region of parameter space come from the UA2 experiment \cite{Alitti:1993pn} but are always much weaker than the monojet constraint. We therefore do not include them in Fig.~\ref{fig:Zprimeconstraints}. When combined with the BBN bound, one sees that the monojet searches exclude $m_{X} \gtrsim 200$~GeV.

A second set of bounds in this mass regime arises from loop-induced kinetic mixing between the $X$ and the standard model photon and $Z$ boson. (Since the twin hypercharge is not gauged, there is no kinetic mixing with the twin $Z$ boson.) In this category there are constraints from electroweak precision measurements, dilepton resonances and exotic decays of $B$ mesons. The effective mixing parameter $\epsilon$, defined by $\mathcal{L}\supset\epsilon X^{\mu\nu}B_{\mu\nu}$, arises from loops with standard model fermions, viz.
\begin{equation}\label{mixingparameter}
\epsilon = - \frac{N_c}{4\pi^2} g_1 g_X\sum_{f} q_f Y_f\int_{0}^1\!\!dx\,x (1-x)\log\frac{x(1-x)m_{X}^2+m_f^2}{x(1-x)\Lambda^2+m_f^2}\,,
\end{equation}
assuming $\epsilon = 0$ at the twin Higgs cutoff scale, $\Lambda = 5$~TeV. Here $g_1$ is the SM hypercharge coupling, and $Y_f$ and $q_f$ are the hypercharge and $U(1)_X$ charge of the fermion $f$ respectively.\footnote{Our definition of $\epsilon$ differs by a factor of $2\cos \theta_w$ from the more conventional normalization (see for instance \cite{Curtin:2014cca}).}  In the universal $X$ coupling case, in which $X$ couples to the third generation, this can be approximated by
\begin{equation}
\epsilon \simeq - \frac{N_c}{24\pi^2} g_1 g_X \bigg[\frac{7}{3}\log\frac{m_{X}^2}{\Lambda^2}+\frac{1}{9}\log\frac{m_{t}^2}{\Lambda^2}\bigg]\simeq g_X\bigg[0.14-0.02\log\left(\frac{m_{X}}{10\;\mathrm{GeV}}\right)\bigg]\,,
\end{equation}
while the analogous expression for the horizontal case is 
\begin{equation}
\epsilon \simeq - \frac{N_c}{12\pi^2} g_1 g_X \log\frac{m_{X}^2}{Q^2}\simeq g_X\bigg[0.11-0.02\log\left(\frac{m_{X}}{10\;\mathrm{GeV}}\right)\bigg]\,.
\end{equation}
As a result of this kinetic mixing, the SM leptons are millicharged under the $X$. Note that because of the tree-level couplings of $X$ to SM and twin quarks, the $X$ branching ratios to jets or missing energy nevertheless dominate leptonic branching ratios when $m_{X}>2m_{\pi}$.

The consequent electroweak precision (EWPT) bounds in Fig.~\ref{fig:Zprimeconstraints} were rescaled from those obtained in Ref.~\cite{Curtin:2014cca}. The dilepton constraints below $m_Z$ are rescaled from those in Ref.~\cite{Hoenig:2014dsa}, where we use Madgraph 5 \cite{Alwall:2011uj} and Feynrules 2.0 \cite{Alloul:2013bka} to compute the cross sections. The projected bound, shown by the dashed orange line, assumes $3$ ab$^{-1}$ of $14$ TeV data. The dilepton bound above $m_Z$ is obtained from a similar recast of the CMS search for $A^0\rightarrow \mu^+\mu^-$~\cite{CMS-PAS-HIG-12-011}. Since the $X$ signal does not contain any $b$-tags, we only make use of `category 3' events in Ref.~\cite{CMS-PAS-HIG-12-011}, which corresponds to gluon-fusion production. We further implement a correction factor of 0.7 to account for the difference in acceptance between this search and the $X$ signal. Given that the branching ratio of the $X$ to leptons is very small, there is no meaningful bound from $h\rightarrow Z X\rightarrow 4\ell$ \cite{Curtin:2014cca}.

There are moreover several applicable $B$ factory searches, that probe this effective kinetic mixing. Firstly, we rescale the result for the search for $e^+e^-\rightarrow X \gamma$ that was carried out in Ref.~\cite{Essig:2013vha}. The 95\% CL excluded region is shown in brown in Fig.~\ref{fig:Zprimeconstraints}. The dashed brown line indicates the expected exclusion reach from Belle II, also rescaled from Ref.~\cite{Essig:2013vha}, using a projected $50$ ab$^{-1}$ of data at the $\Upsilon(4S)$ resonance.  Secondly, we adapt the branching ratio $\mathcal{B}(\Upsilon(1S)\to\text{invisible})<3\times 10^{-4}$ at $90\%$ CL from BaBar \cite{Aubert:2009ae}. In the universal coupling case, this can be interpreted as a bound on the invisible decay $\Upsilon(1S)\rightarrow X^\ast \rightarrow q'q'$ decay, since for $m_{X}\geq 2m_{\pi}$ and $g_X\lesssim0.1$ the twin hadrons produced in the $X$ decay all decay outside the detector. We extract a bound on $g_X$ from Refs.~\cite{Yeghiyan:2009xc,Graesser:2011vj} by comparing the invisible branching ratio with the well-measured branching ratio of $\Upsilon(1S)\to\mu^{+}\mu^-$,
\begin{equation}
\label{eqn:B2BRC}
\frac{\mathcal{B}(\Upsilon(1S)\to q'q')}{\mathcal{B}(\Upsilon(1S)\to\mu^{+}\mu^-)}\approx\frac{3}{2}\left(\frac{g_X}{e^2}\frac{m_{\Upsilon}^2}{m_{X}^2-m_{\Upsilon}^2}\right)^2<1.2\times10^{-2}\,,
\end{equation} 
where we assumed that the unknown hadronic matrix elements for each process are approximately the same. The result is shown in the purple-shaded region in Fig.~\ref{fig:Zprimeconstraints}. The projected sensitivity of Belle II is indicated by the dashed purple line, and is derived by assuming the branching ratio constraint \eqref{eqn:B2BRC} will improve to the $\simeq 4\times 10^{-4}$ level, as discussed in Ref.~\cite{BelleII}. Finally, for the universal case it is plausible that a slightly stronger bound can be obtained by recasting the BaBar search for $\Upsilon(3S)\to\gamma+$pseudoscalar \cite{Lees:2011wb}. However, in this case a careful treatment of the hadronic matrix element is necessary, which is beyond the scope of this paper. (See for instance Ref.~\cite{Yeghiyan:2009xc} for an analysis in terms SM-DM effective operators.)

We see in Fig.~\ref{fig:Zprimeconstraints} that the projected reach of Belle II dark photon searches and dilepton searches should probe nearly all of the presently allowed parameter space.

$\bm{m_{X} < 1~\mbox{GeV}}$: In this regime a different set of constraints becomes relevant. First, the EFT approximation leading to the BBN bound \eqref{eqn:BBNXM} is no longer valid, and the $X$ must be included as a dynamical degree of freedom. This does not significantly affect the bound, except for $m_{\twin{\pi}^0}$ nearby to the $X$ resonance or if $m_{X}<m_{\twin{\pi}^0}/2$. In this latter case the twin pion can directly decay to two on-shell $X$ bosons through the $U(1)_X$ chiral anomaly, rather than a diphoton final state, and the BBN bound no longer depends on the $X$ mass.

\begin{figure}[t]
	\includegraphics[width=0.49\textwidth]{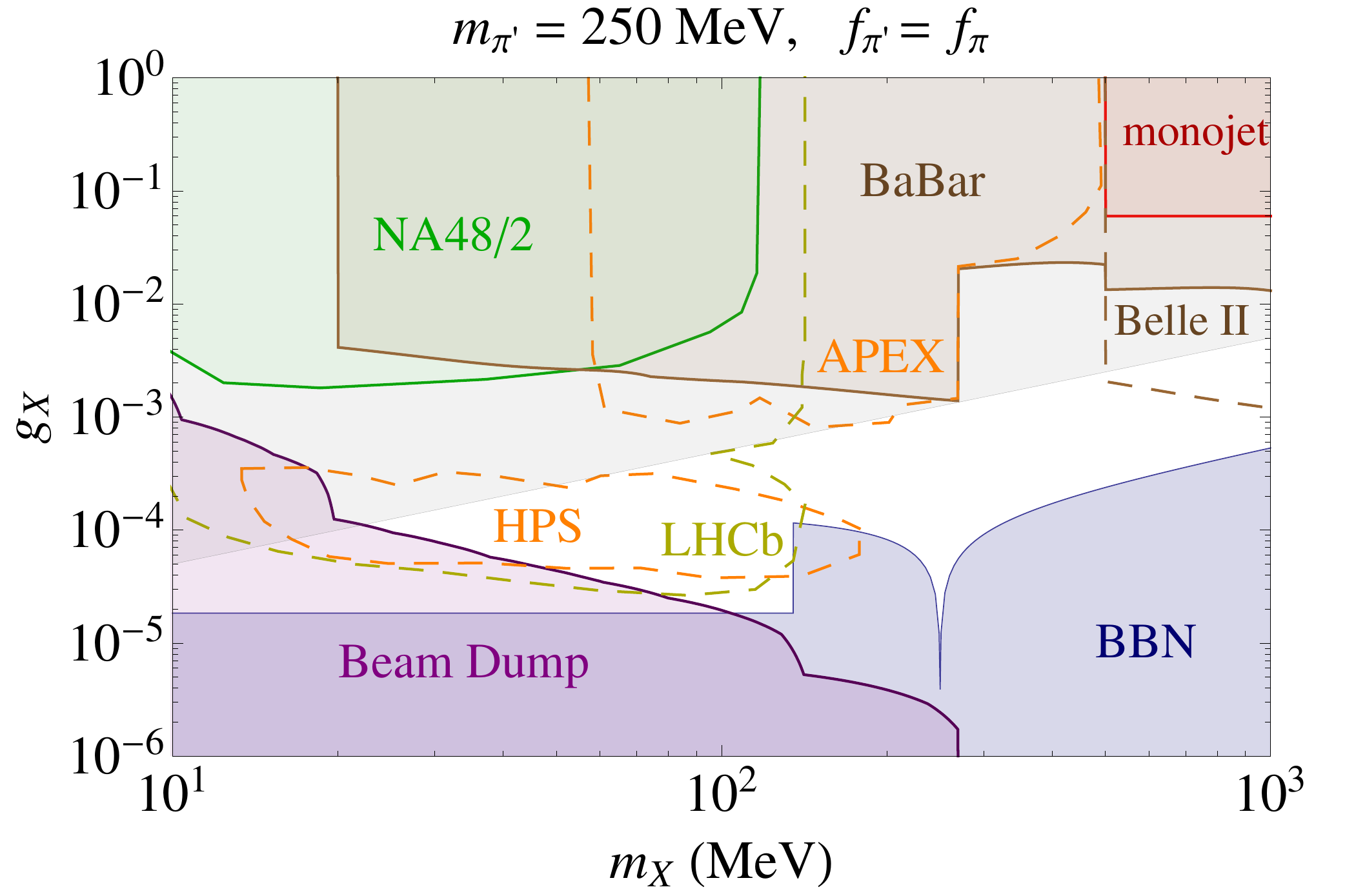}\hfill
	\includegraphics[width=0.49\textwidth]{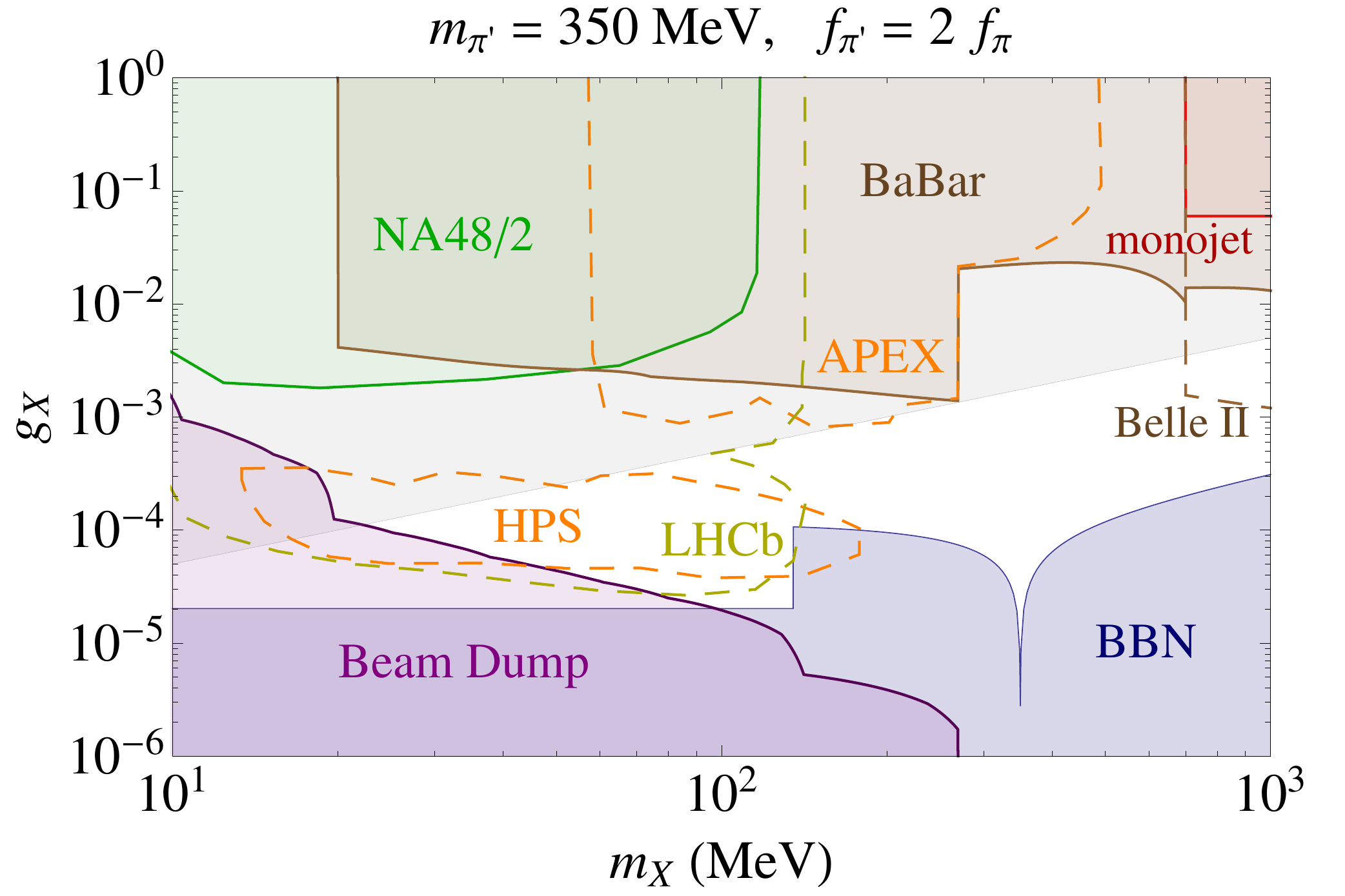}
	\caption{Constraints on the $m_{X}$--$g_X$ parameter space for both benchmarks \eqref{eqn:MFB}, for the universal model only. See the text for a detailed discussion of the constraints. Dashed lines correspond to projections of future experimental reach. In the gray shaded region, at least 10$\%$ fine-tuning is required to keep the $X$ light. The constraints for the horizontal model are nearly identical, except for the absence of the tuning constraint (gray shaded area).}
	\label{fig:ZprimeconstraintsLow}
\end{figure}

Second, since $\alpha_s$ is large in this regime, one expects $\mathcal{O}(1)$ incalculable corrections to the mixing \eqref{mixingparameter}, which increases the theoretical uncertainty on the bounds that rely on kinetic mixing. Here we continue to use eq.~\eqref{mixingparameter} for definitiveness, while keeping these $\mathcal{O}(1)$ uncertainties in mind.  In Fig.~\ref{fig:ZprimeconstraintsLow} we show all applicable constraints in this regime for the universal model. The horizontal model is nearly identical. The BaBar and monojet bounds, indicated by the brown and red shaded region respectively, are sensitive to various thresholds involving SM and twin pion masses. Specifically, once $m_{X}< 2 m_{\pi'}$, the $X$ can no longer decay to the twin sector. As a result both the monojet bound and the BaBar bound on $\gamma$+MET disappear. In this regime the most powerful BaBar bound comes from $X\rightarrow\ell^+\ell^-$ decay \cite{Lees:2014xha}, which begins to dominate once the decay to SM pions is forbidden, i.e. for $m_{X}< 2 m_{\pi}$. 

Finally, for very a light $X$ there are constraints from $\pi\rightarrow X\gamma$ decay at the NA48/2 experiment \cite{Goudzovski:2014rwa}, as well as from beam dump experiments \cite{Andreas:2012mt}. For completeness we also include projected bounds from APEX \cite{Schuster:2013sma} and HPS \cite{Moreno:2013mja}, in addition to a projected bound from $D^{\ast 0}\rightarrow D^0 X$ at LHCb \cite{Ilten:2015hya}. Since the $X$ couples directly to the quark sector, the rate in NA48/2  and LHCb receives a contribution in addition to that from the kinetic mixing. Since this contribution comes with an incalculable hadronic matrix element, we simply take it to be zero here, which is very conservative. 

In summary, the constraints are qualitatively similar for the universal and horizontal cases, and that much of the remaining available parameter space will be probed in upcoming experiments.

\section{Gamma-ray spectra}
\label{sec:indirect}

Since the $\twin{\pi}^0$s decay mostly to photons -- diphoton final states for $m_{\twin{\pi}^0} < 3m_{\pi}$, and higher multiplicity soft photons for $3m_{\pi} < m_{\twin{\pi}^0} < 2m_K$ -- the dark shower of light twin hadrons produced by the DM annihilation can result in an astrophysically detectable signal. Compared to stable (or long-lived) charged decay products, the production spectra and distribution of photons do not suffer from large corrections arising from propagation, and therefore provide robust probes or constraints on the underlying DM distribution and annihilation process. 

Over the last decade, much attention has been focused on a galactic center $\gamma$-ray excess (GCE), claimed to be seen in \emph{Fermi}-LAT data from the central regions of the Milky Way galaxy~\cite{Goodenough:2009gk,Hooper:2010mq,Hooper:2011og,Hooper:2013rwa,Abazajian:2014fta,Daylan:2014rsa,Calore:2014nla,Calore:2014xka,TheFermi-LAT:2015kwa}  that may be a signal of DM annihilations into a variety of SM final states. Corresponding excesses in dwarf spheroidal galaxies are also claimed to be seen~\cite{Geringer-Sameth:2015lua,Hooper:2015ula}, though this is disputed by the \emph{Fermi} collaboration \cite{Drlica-Wagner:2015xua}. The observed GCE flux corresponds to an underlying DM annihilation rate comparable to a thermal relic cross section, and exhibits a spherically symmetric morphology consistent with a NFW-type DM profile. Although the DM explanation of the GCE has been called into question~\cite{Abazajian:2012dgr,Abazajian:2012edg,Gordon:2013vta,Cholis:2014lta,Cholis:2014noa,Lee:2014mza,Lee:2015fea}, and should be approached with caution, models with DM annihilations at rates comparable to thermal relic cross-sections can generically produce photon fluxes comparable to current experimental sensitivities of the \emph{Fermi}-LAT. Hence, any thermal DM model that predicts $\gamma$-ray signals from annihilation needs to be consistent with the observed $\gamma$-ray spectra, no matter if treated as a signal or constraint. 

Cosmic rays -- antiprotons and positions -- are typically produced alongside $\gamma$-ray spectra in many frameworks designed to address the GCE, possibly resulting in mild tensions with Pamela and AMS-02 data (see e.g.~\cite{Bergstrom:2013jra,Bringmann:2014lpa,Bringmann:2009ca,Cirelli:2014lwa,Pettorino:2014sua,Cholis:2014fja}). However, DM annihilation into cascade processes provides a generic way to soften the spectra of cosmic rays, and hence loosens the corresponding astrophysical bounds~\cite{Elor:2015tva}. Most previous analyses have focused on cascades with fixed multiplicity, but another option is to consider showering in a strongly-coupled dark sector \cite{Freytsis:2014sua}. This is precisely the situation arising from T-WIMP annihilations into showers of light twin hadrons within the hadrosymmetric twin Higgs framework. In this section we adapt our prior treatment in Ref.~\cite{Freytsis:2014sua} of this so-called dark showering to simulate the $\gamma$-ray spectra produced by such DM annihilations.

The spectra are generated by \textsc{Pythia8}~\cite{Sjostrand:2006za}, with the QCD parton shower and hadronization model modified appropriately for the twin quark sector. In detail, in order to be able to correctly approximate the twin shower spectrum, the running of \textsc{Pythia8} has to be suitably generalized to allow for the variation of twin quark masses relative to the twin confinement scale. This involves two considerations: First, the mass thresholds for the running of strong coupling and the parton shower cutoff are adjusted relative to the quark masses to make sure that hadronization into the heaviest pNGBs of the theory  -- the analogs of the $K$ and $\eta$ mesons -- is still kinematically allowed. Second, the spectra of hadrons has to be appropriately shifted with the twin quark masses. The full list of QCD hadrons in \textsc{Pythia8} is extensive, and has many parameters extracted from data that are not easily explained from first principles. For the twin shower, only the two lowest lying $SU(3)$ flavor multiplets for the mesons and baryons were retained. To validate this approach, it was checked that when the parameters of the confining gauge group matched those of QCD, this did not introduce appreciable differences between the pion and analogous twin pion energy distributions. 

Of the light quark twin hadrons, all but the pNGB octet masses were taken to scale proportionally with $\twin{\LambdaQ}$ just as in eq.~\eqref{eqn:THM}, while the pNGB masses are scaled as in eq.~\eqref{eqn:TPM}.  Combined with eq.~\eqref{eqn:TQM}, this implies the scaling relation
\begin{equation}
\label{eqn:QPLS}
\frac{m_{\twin{q}}}{m_q} = \left(\frac{m_{\twin{\pi}}}{m_{\pi}}\right)^2 \frac{\LambdaQ}{\twin{\LambdaQ}}\,.
\end{equation}
For the purpose of these simulations, we take the SM confinement scale $\LambdaQ \simeq 250$~GeV.

The simulations are valid in both the two and three light flavor regimes, as in eq.~\eqref{eqn:LQCDB}. Interpolation between the three and two light flavor regions of parameter space requires a change in the mass scaling relations of the twin hadrons. The precise location of transition from one regime to the other is not well-defined, and no heutristics for the spectrum of particles with $m_{\twin{q}} \sim \twin{\LambdaQ}$ exist. In all figures in this section, we demarcate the three light flavor regime with grey dashed lines, according to the definition in eq.~\eqref{eqn:LQCDB}, and as shown in Fig.~\ref{fig:LR}. The transition region to the two light flavor theory beyond this demarcation should be taken as capturing the rough dependence of the spectra on the parameters. Parametrically far away from the boundary in either direction our simulations should be considered more quantitatively reliable. Once the twin strange quark can be considered heavy, the masses of all pseudoscalar mesons with significant strange content  -- the twin $K$ and $\eta$ mesons -- are taken to scale linearly with the quark mass. The same becomes true of the vector meson and baryon multiplets with twin strange content, that now scale with $m_{\twin{s}}$ rather than $\twin{\LambdaQ}$. However,  these heavier states have a less significant effect on the spectrum of the dark shower.

Computation of astrophysical rates is done using \textsc{PPPC 4 DM ID}~\cite{Cirelli:2010xx}. With this setup, we proceed to assess the compatibility of our DM annihilation signals with the observed Milky Way galactic center $\gamma$-ray spectrum, whether treated as an actual spectrum to be fit, or as a possible background providing a constraint.

\subsection{GCE fits}
\label{sec:GCEfit}

\begin{figure}[t]
	\includegraphics[width=0.45\textwidth]{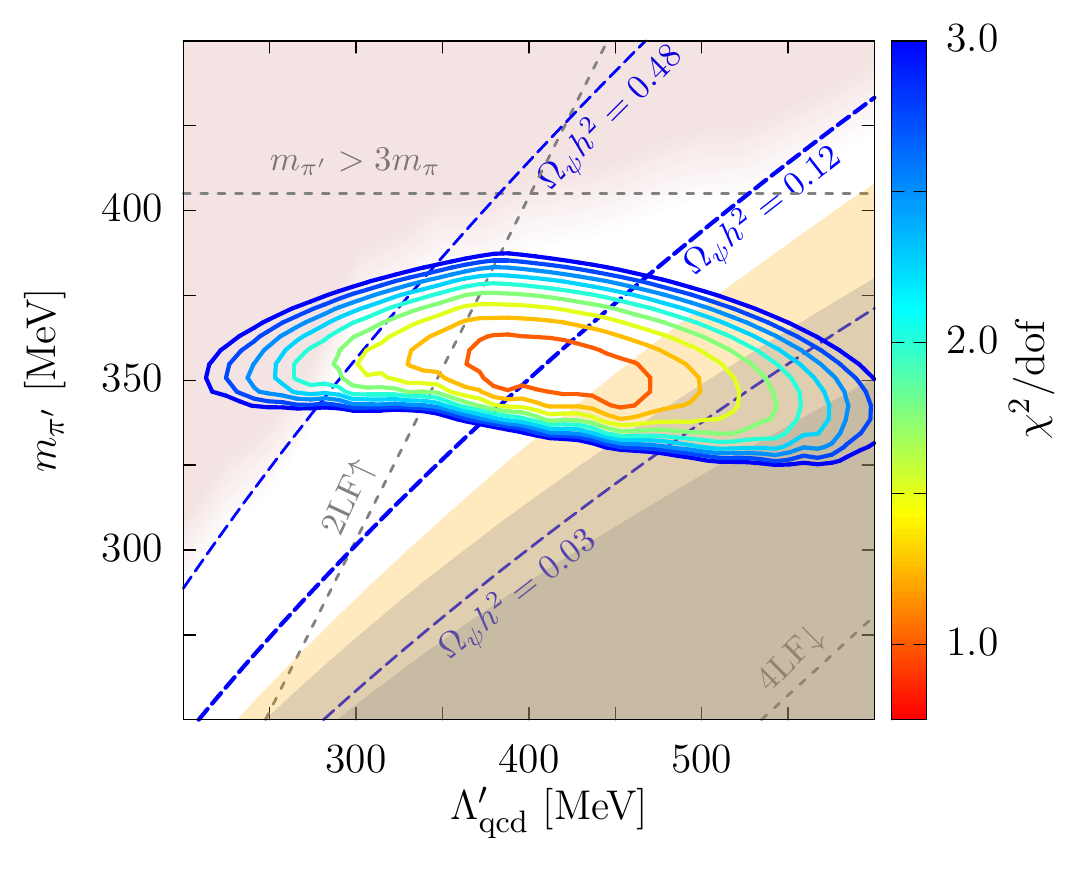}\hfill
	\includegraphics[width=0.45\textwidth]{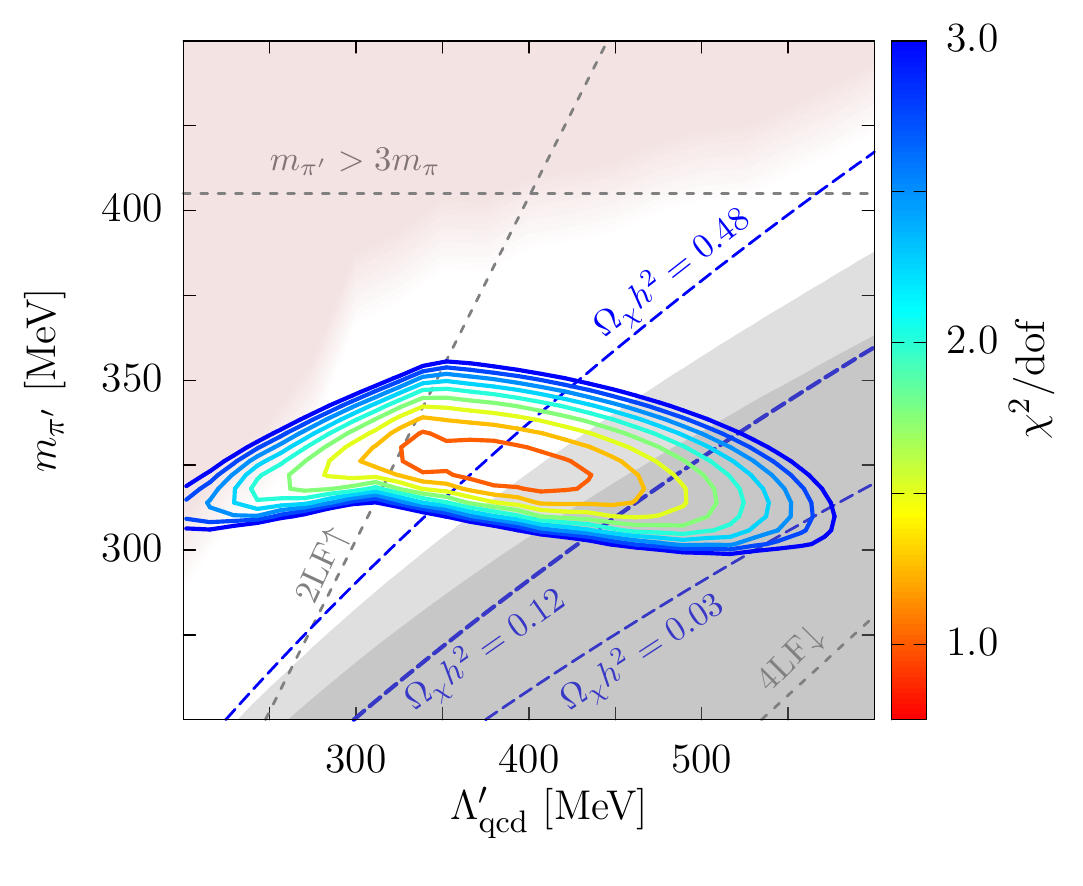}
	\includegraphics[width=0.45\textwidth]{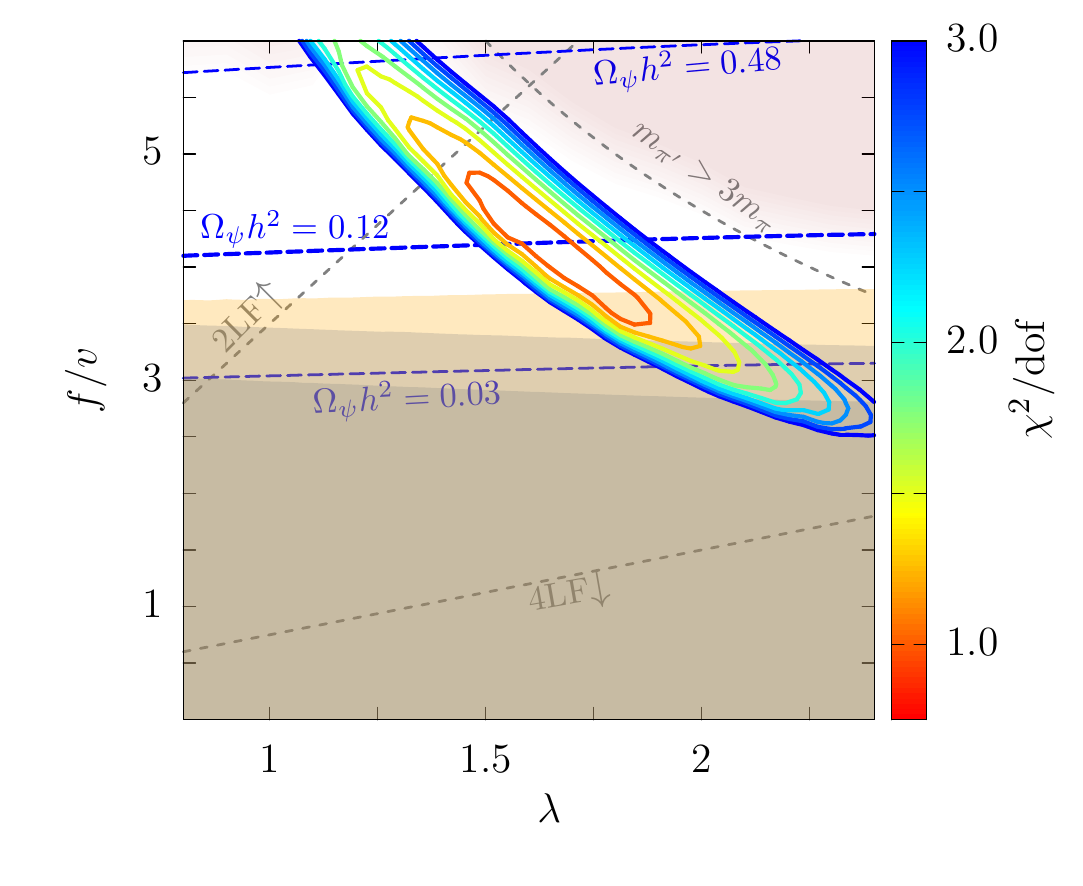}\hfill
	\includegraphics[width=0.45\textwidth]{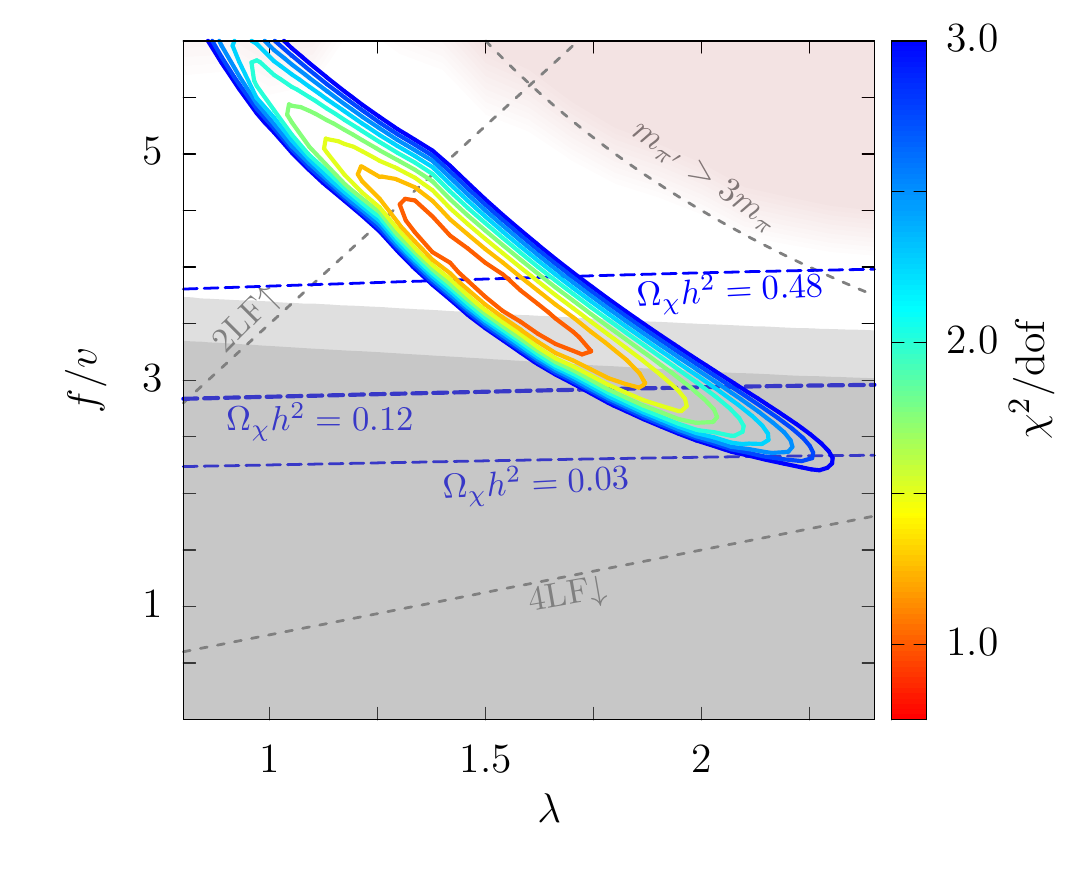}
	\caption{Goodness-of-fit ($\chi^2/\textrm{dof}$) contours for the GCE spectrum for the Dirac (left) and Majorana (right) scenarios, shown in $m_{\twin{\pi}}$--$\twin{\LambdaQ}$ parameter space. Equivalent plots are shown below for the $f/v$--$\lambda$ parametrization. At each point in parameter space, the DM mass is floated to find the best fit to the spectrum. Dashed grey lines indicate the boundaries of the three light flavor and $m_{\twin{\pi}} < 3m_{\pi}$ region (cf. Fig.~\ref{fig:LR}).  Contours of constant relic abundance, determined from eqs.~\eqref{eqn:FVDM}, are denoted by heavy (light) dashed blue lines for $100\%$ ($400\%$ and $25\%$) DM abundances.  Also shown are Higgs invisible width exclusion regions with and without $b'\bar{b'}$ decay modes (light grey and dark grey, respectively), as well as CMB reionization bounds (orange). See Fig.~\ref{fig:RI} for details. Red shading indicates regions of large sensitivity to the details of the twin mass hadron spectrum (see text for more details). Spectra provided by Ref.~\cite{Calore:2014xka}.}
	\label{fig:GCEfit}
\end{figure}

In order to understand the effect of the twin parton shower on the $\gamma$-ray spectrum, we simulated spectra for different choices of DM mass, $m_{\textrm{DM}}$, the Landau pole of the twin confining gauge group, $\twin{\LambdaQ}$, and twin quark masses, $m_{\twin{q}}$. We parametrize the latter by the mass of $\twin{\pi}^0$, via the relation \eqref{eqn:QPLS}. The features of the $\gamma$-ray spectrum are then fully characterized by either $m_{\textrm{DM}}$, $\twin{\LambdaQ}$ and $m_{\twin{\pi}}$ or equivalently by $m_{\textrm{DM}}$, $f/v$ and $\lambda$, via the scalings \eqref{eqn:LQR} and \eqref{eqn:TPM}.

\begin{figure}[t]
	\centering\includegraphics[width=0.45\textwidth]{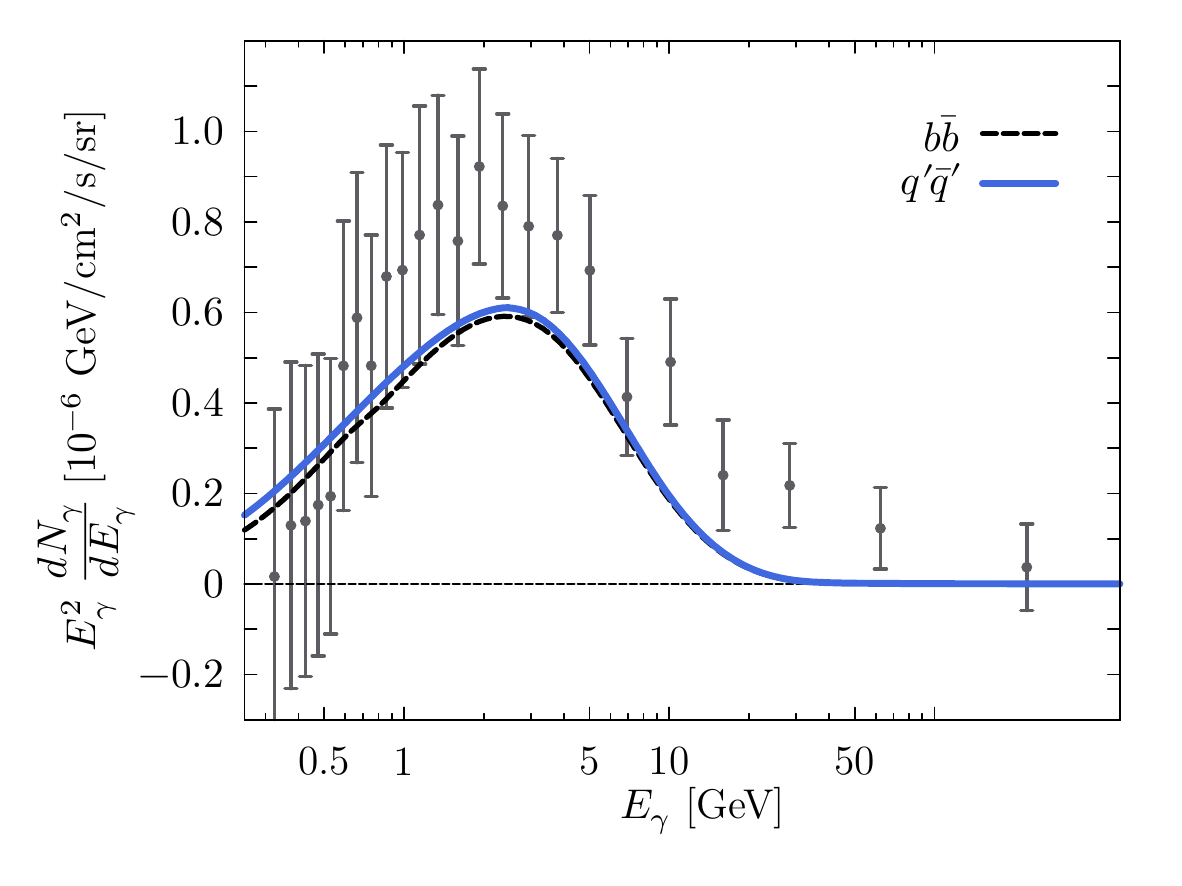}\hspace{0.06\textwidth}
	\caption{The overall best-fit photon spectrum for the GCE, arising from DM annihilation to twin quark states and subsequent dark showering. Also shown is the best fit spectrum for annihilation directly to $b\bar{b}$ final states, which provides the best fit when considering only direct annihilation to 2-body SM states. The goodness-of-fit in the two cases is essentially identical, being $\chi^2/\text{dof} = 24.45/24$ and $\chi^2/\text{dof}_{b\bar{b}} = 25.83/24$, respectively. Only the self-variance in each energy bin is displayed, while the fits are performed taking into account correlations between energy bins. The resulting best-fit spectra therefore differ from those expected from a visual fit. See text for details. Spectra extracted from Ref.~\cite{Calore:2014xka}.}
	\label{fig:GCEspec}
\end{figure}

Goodness-of-fit ($\chi^2/\textrm{dof}$) contours for fits of photon spectra produced by T-WIMP annihilations to the GCE signal are shown in Fig.~\ref{fig:GCEfit}, both for $m_{\twin{\pi}}$--$\twin{\LambdaQ}$ and $f/v$--$\lambda$ parameter spaces, while the spectrum at the point of best fit is shown in Fig.~\ref{fig:GCEspec}, and compared with the best fit for direct annihilation into $b\bar{b}$ pairs. Because of correlations between the energy bins~\cite{Calore:2014xka}, the best fit curves differ from those that would be expected under the assumption of purely uncorrelated signal bins. Namely, while the overall normalization is dominated by the lowest energy bins, the shape of the spectrum is predominately determined by the peak and the beginning of the tail region. This results in a fit that appears to anomalously undershoot the highest-energy bins.

Since the uniformity of the scaling relations \eqref{eqn:THM} and \eqref{eqn:TPM} should only be taken as approximate, we have checked to see the effect of varying the masses of the excited multiplets by an additional 10\% around their nominal values. As twin quark masses relative to the confinement scale increase, this variation introduces larger and larger deviations in the spectrum. In Fig.~\ref{fig:GCEfit}, we shade in red the region where these deviations change the $\chi^2/\text{dof}$ goodness-of-fit metric by more than unity, which we take as a sign that the shower is sensitive to the non-perturbative aspects of the hadron spectrum at a level that makes spectrum fits unreliable. For fixed $x_f (= 20)$, the DM relic abundance \eqref{eqn:FVDM} is uniquely determined by $f/v$ and $m_{\textrm{DM}}$.  Since in turn $\lambda f/v =  (m_{\twin{\pi}}/m_{\pi})^2$ and $\lambda = \twin{\LambdaQ}/\LambdaQ$, from eqs.~\eqref{eqn:TPM} and \eqref{eqn:LQR}, the relic abundance may be reparameterized in terms of $\twin{\LambdaQ}$ and $m_{\twin{\pi}}$. The consequent relic abundance `theory contours' are also displayed in Fig.~\ref{fig:GCEfit}.

The fitted DM masses range only from $56$ to $59$~GeV over the entire displayed parameter space for both scenarios, with the Dirac and Majorana DM best-fit masses both being $\simeq 57$~GeV. In both cases, the best GCE fit regions are partially excluded by Higgs invisible width and CMB reionization bounds. For the fits shown in Fig.~\ref{fig:GCEfit}, we can see that the best fit regions for the spectra correspond to a twin confinement scale slightly higher than that of the SM, and the twin quark mass to confinement ratio $m_{\twin{q}}/\twin{\LambdaQ} = (f/v)/\lambda (m_q/\LambdaQ)$ is slightly higher than in the SM sector, too. Moreover, the best fit region lies in the three light flavor regime. With respect to the $\Omega h^2$ theory contours, we see that the remaining allowed region for the Majorana scenario is overclosed, indicating that the Majorana DM scenario is excluded as an explanation for the GCE. In the case of Dirac DM, however, the theory contour corresponding to the observed DM relic abundance, $\Omega_\psi h^2 = 0.12$, passes very close to and through the non-excluded region of best GCE fit. Further, one sees that this contour as well as the best-fit regions have $f/v \simeq 4$, which corresponds to a fine-tuning of only $\mathcal{O}(10\%)$. This is close to the minimal fine-tuning permitted by current bounds. 

\begin{figure}[t]
	\includegraphics[width=0.45\textwidth]{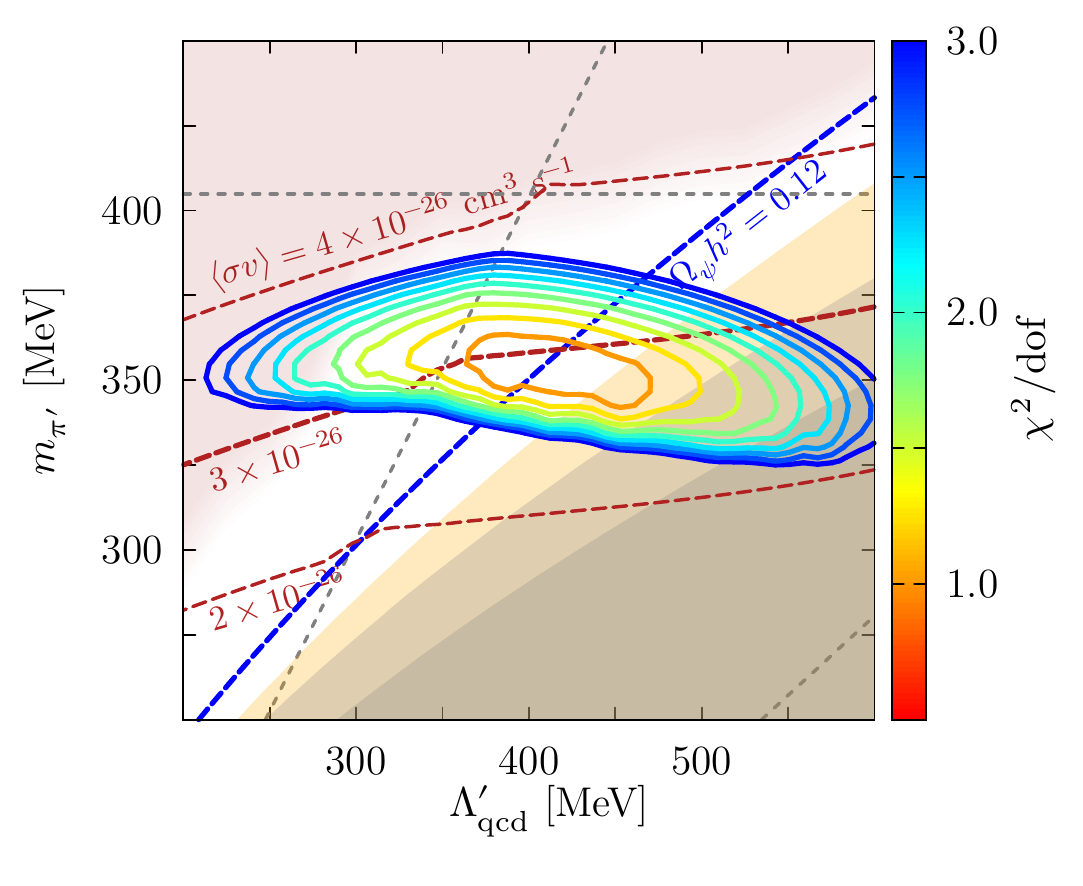}\hfill
	\includegraphics[width=0.45\textwidth]{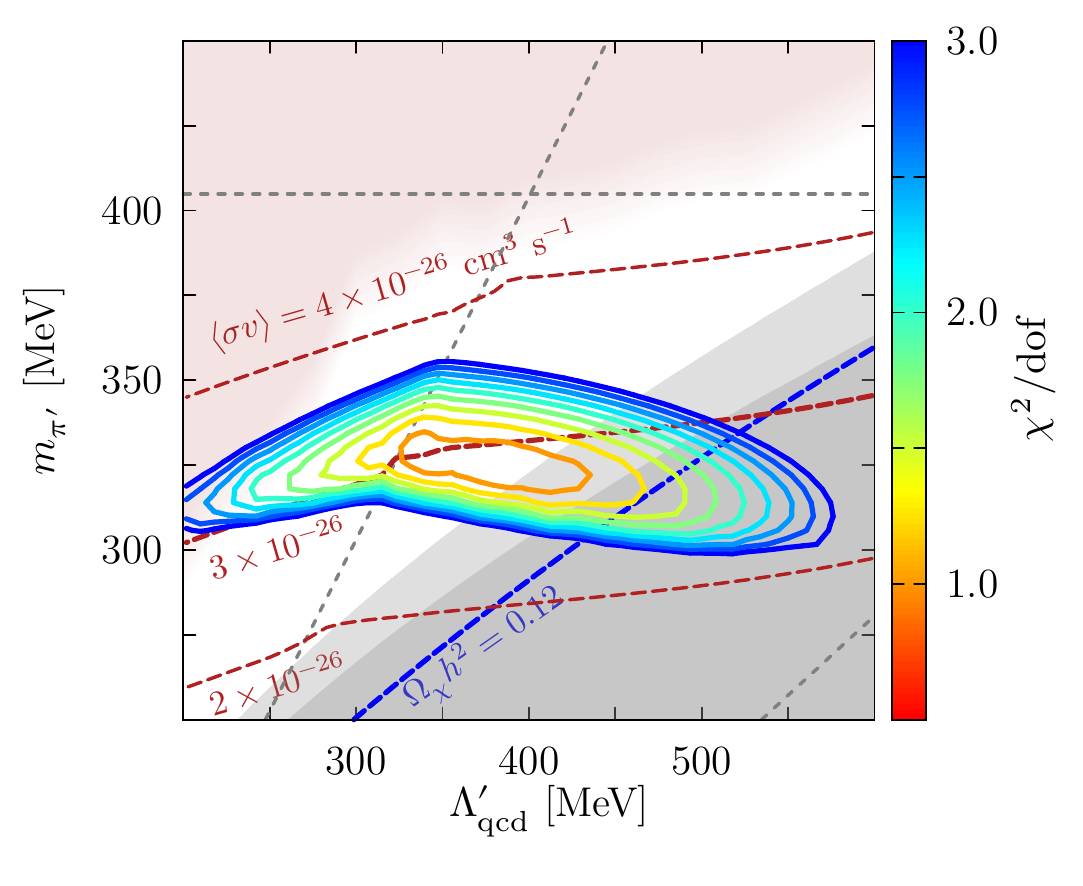}
	\caption{Contours of constant DM annihilation cross-section, as determined from the GCE fits, overlaid upon goodness-of-fit ($\chi^2/\textrm{dof}$) contours for the GCE spectrum for the Dirac (left) and Majorana (right) scenarios. Exclusion regions are shown just as for Fig.~\ref{fig:GCEfit}. The annihilation cross-section contours are indicated by heavy (light) dashed blue lines for $\langle \sigma v \rangle = 2$, $3$ and $4\times 10^{-26}$~cm$^{3}$s$^{-1}$. Spectra extracted from Ref.~\cite{Calore:2014xka}.}
	\label{fig:GCEfit2}
\end{figure}

In Fig.~\ref{fig:GCEfit2} we show the GCE goodness-of-fit contours together with contours of constant DM annihilation cross-section as extracted from the GCE spectral fits, rather than from the relic abundance theory contours  \eqref{eqn:FVDM}. In the Dirac case, the annihilation cross-section contour corresponding to the observed DM relic density, $\langle \sigma v \rangle \simeq 3\times 10^{-26}$~cm$^{3}$s$^{-1}$, intersects the best fit GCE region, which is by itself a non-trivial agreement. We see further that this annihilation cross-section contour moreover intersects the $\Omega_\psi h^2 = 0.12$ theory contour in the best fit region: A remarkable concordance between predictions of the hadrosymmetric twin Higgs framework and the observed GCE spectral features.

Besides GCE signals, DM capture and subsequent annihilation in the Sun might also provide detectable $\gamma$-ray spectra. In the Dirac scenario, $\sim 60$ GeV DM particles have a nuclear scattering cross-section $\sigma_N\simeq 10^{-46}\text{cm}$, which corresponds to a capture rate $C_{\odot}\simeq 10^{19}\,\text{s}^{-1}$ \cite{Schuster:2009fc, Garcia:2015loa}. This is sufficiently fast to ensure that equilibrium will be reached, which implies a DM annihilation rate $\Gamma_{\text{ann}}=\frac{1}{2}C_{\odot}$. If twin pions produced in the dark shower have a typical lifetime $\tau \lesssim 1$~s, most of them escape from the solar interior before decaying to $\sim$ GeV diphotons that may be seen by terrestrial experiments. For $m_{\twin{\pi}} \simeq 200$ MeV, the incident photon flux is approximately of order $10^{-8}$~cm$^{-2}$s$^{-1}$, which is comparable to the $\gamma$-ray flux seen by the \emph{Fermi}-LAT quiescent sun data \cite{2011ApJ...734..116A}, or the flux bound recast from a leptonic final state analysis in Ref.~\cite{Schuster:2009au}. It would therefore be interesting to examine the future sensitivity of solar observations to this DM annihilation signal. We leave a careful study of these solar constraints to future work.

\subsection{Galactic center constraints}

Treating the observed galactic center $\gamma$-ray spectrum as an upper bound on the total photon flux due to DM annihilation plus backgrounds, we may instead construct constraints on the $m_{\textrm{DM}}$--$\twin{\LambdaQ}$ parameter space along the DM relic density theory contour, $\Omega_\psi h^2 = 0.12$. Here we consider a DM annihilation photon spectrum to be consistent with the galactic center data, if the spectrum does not exceed in any bin the total observed photon flux \cite{Calore:2014xka} anywhere in its energy range. Note again that $m_{\twin{\pi}}$ or $f/v$ is fully determined along the $\Omega h^2$ contour for each point in $m_{\textrm{DM}}$--$\twin{\LambdaQ}$ parameter space via eqs.~\eqref{eqn:FVDM}.

\begin{figure}[t]
	\includegraphics[width=0.45\textwidth]{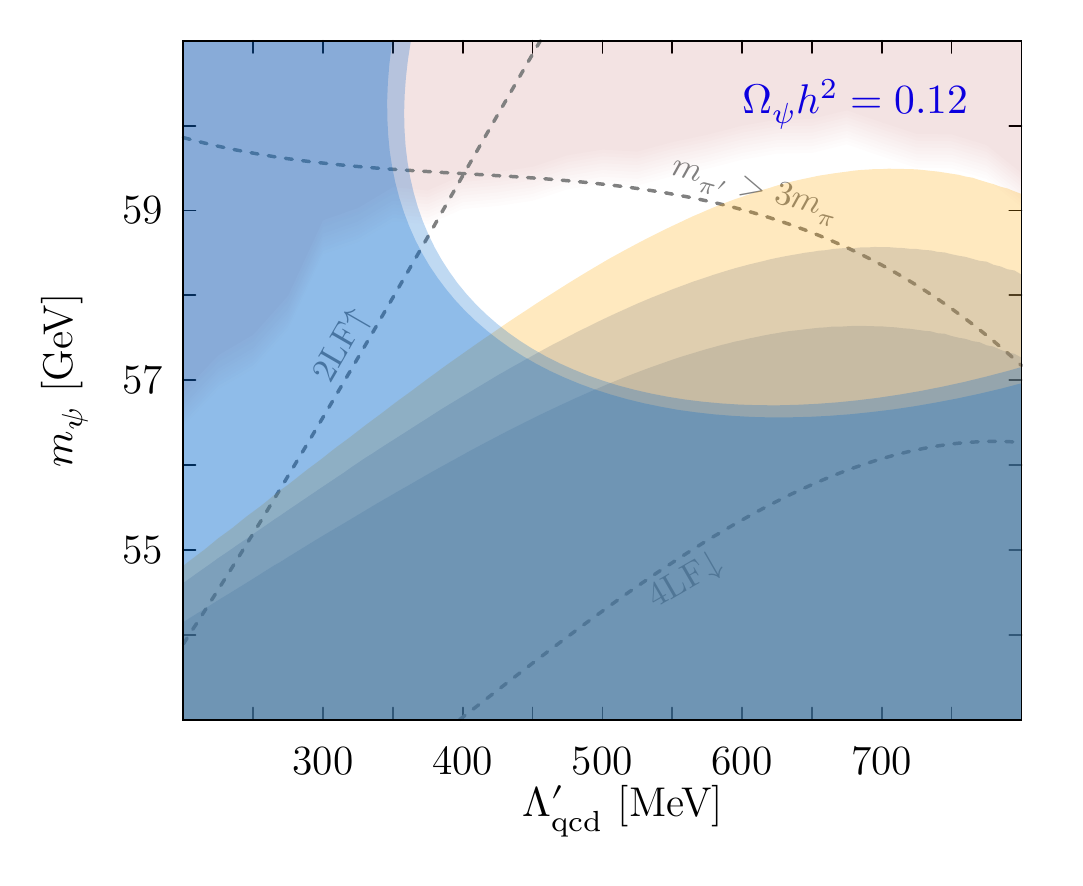}\hfill
	\includegraphics[width=0.45\textwidth]{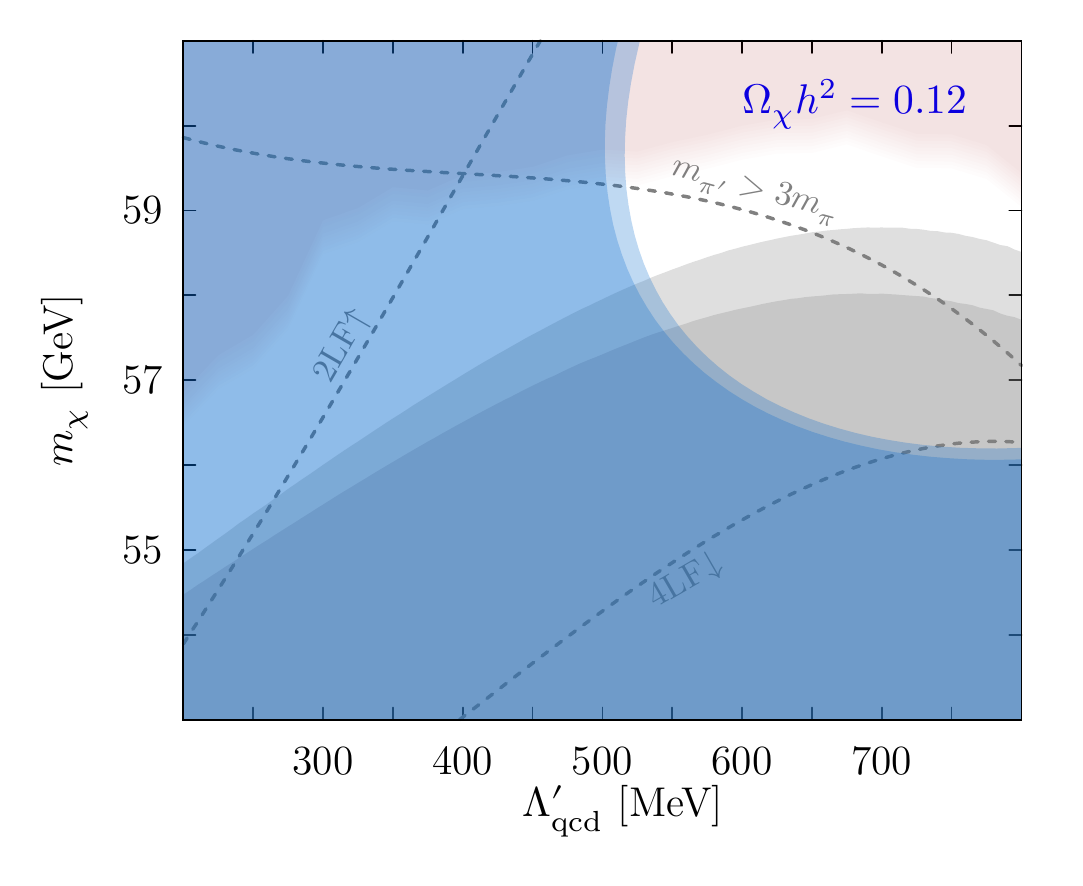}
	\includegraphics[width=0.45\textwidth]{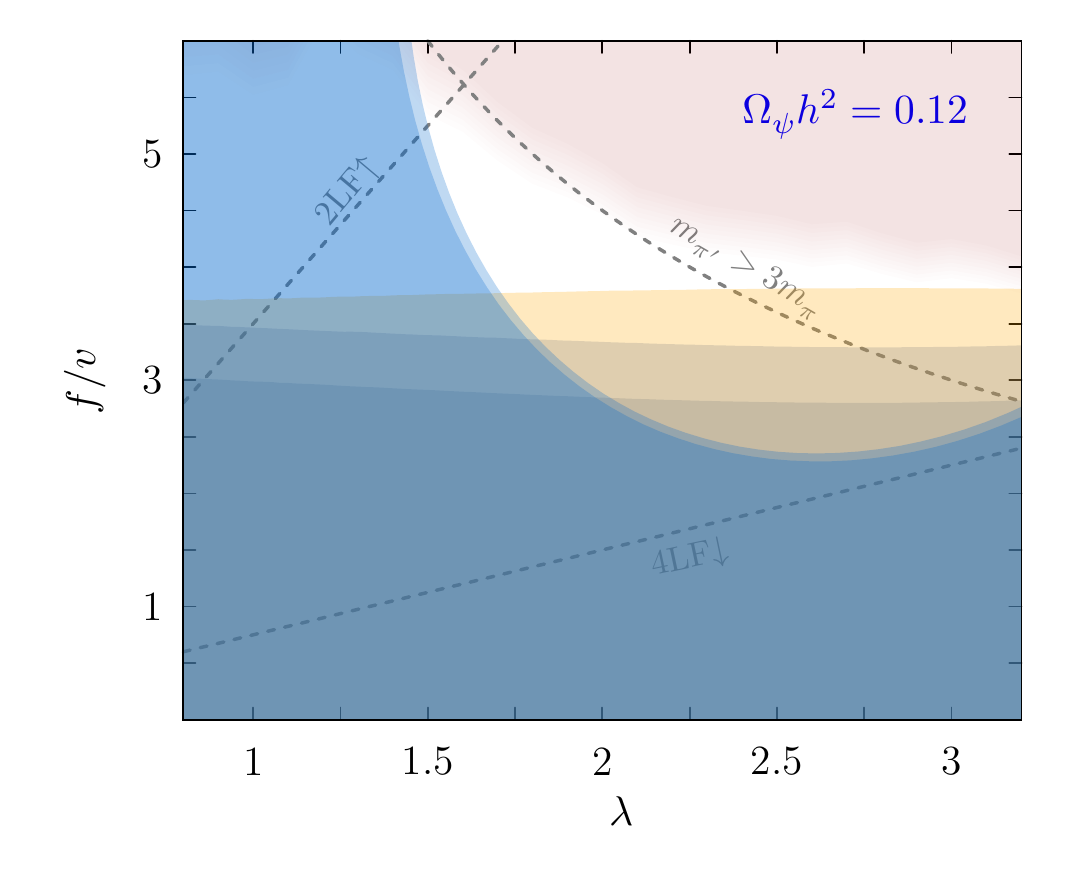}\hfill
	\includegraphics[width=0.45\textwidth]{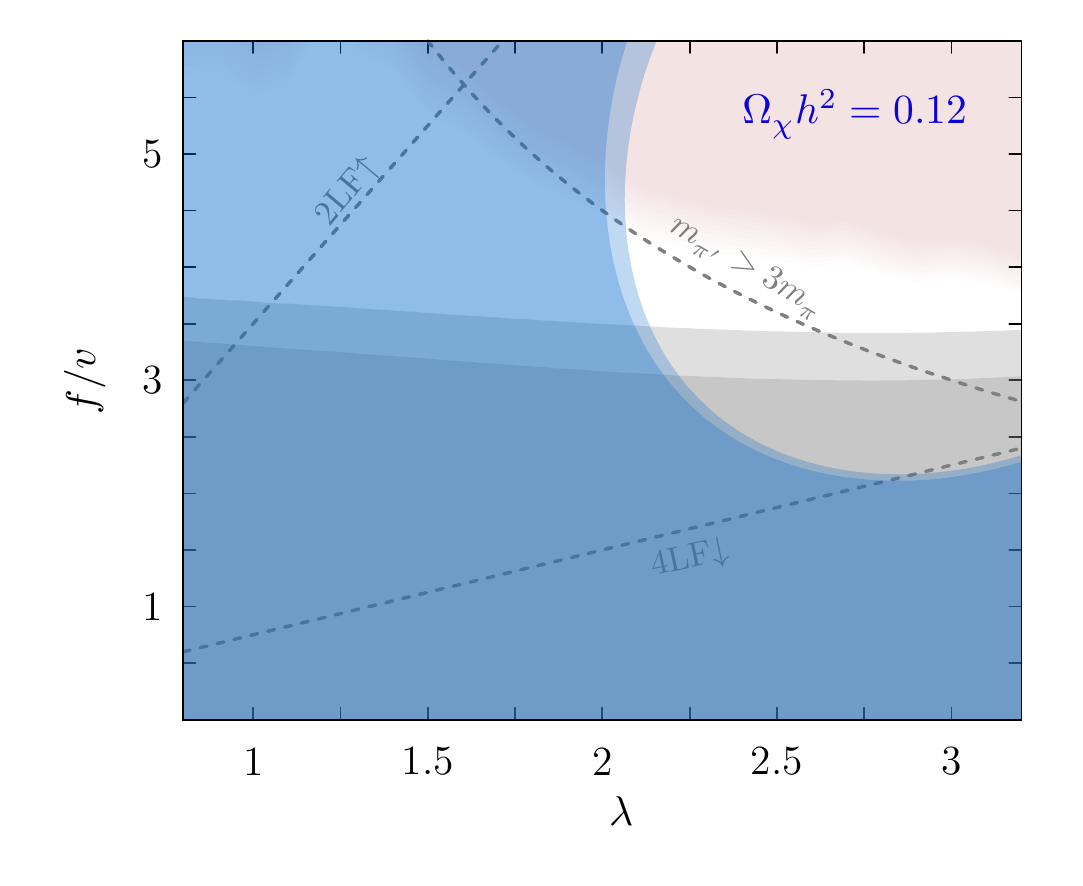}
	\caption{Exclusion regions from current GCE spectrum measurements at 68\% (light blue) and 90\% CL (dark blue) for the Dirac (left) and Majorana (right) scenarios. At all points, the relic density is fixed to $\Omega_\text{DM} h^2 = 0.12$.  Other exclusion regions are shown just as for Fig.~\ref{fig:GCEfit}.  Spectra provided by Ref.~\cite{Calore:2014xka}. }
	\label{fig:GCEcons}
\end{figure}

The galactic center constraints are shown in Fig.~\ref{fig:GCEcons}. Regions of parameter space previously identified in Fig.~\ref{fig:GCEfit} as consistent with the GCE spectrum, including the best fit region, obviously still remain viable. Regions with twin pion mass and confinement scales larger than in the best fit region of Fig.~\ref{fig:GCEfit} feature twin hadron masses that are closer to the DM masses, and therefore their dark showers exhibit a lower multiplicity of hadrons, and hence smaller photon fluxes. In both the Dirac and Majorana cases, such regions remain consistent with the observed $\gamma$-ray spectra. For the purpose of setting constraints, the regions for which modeling of the twin showering spectrum becomes unreliable -- the red shaded regions -- are less consequential than they were in the prior case of carrying out a GCE fit. This is because in regions of parameter space where the $\gamma$-ray spectra are significantly below current sensitivities, even substantial variations in the showering spectrum cannot lead to detectable effects. Further, unless the variations in such spectra push them to the point of detectability, these variations can safely be ignored. As a result, the red shaded regions are smaller here compared to those in Figs~\ref{fig:GCEfit} and \ref{fig:GCEfit2}.

Besides bounds from galactic center data, there are also constraints from the photon fluxes generated by dwarf spheroidal galaxies (dSphs). Some studies (see e.g. Ref.~\cite{Drlica-Wagner:2015xua}) have found significant constraints on various $2$--$2$ DM to SM annihilation channels, that may otherwise explain the GCE. These results, however, depend on estimates for the astrophysical $J$-factors of the relevant DSphs, which require confirmation from observation. Future measurements of the kinematics of the member stars in dSphs can resolve this issue, and may provide additional constraints on the dark showering scenario, beyond those shown in Fig.~\ref{fig:GCEcons}.

Finally, let us consider the constraints from $\gamma$-ray spectra in the case that $m_{\twin{\pi}} > 3m_\pi$ and less than $2m_K$. In this regime, as discussed in Sec.~\ref{sec:DRBRE} above, the dominant twin pion decays are to $\pi^+\pi^-\pi^0$ or $3\pi^0$ final states, such that a higher multiplicity of softer photons are produced by the dark showers, along with soft muons and neutrinos: The photon to lepton production ratio from twin pion decays is expected to be approximately 11\,:\,4 in favor of photons. Since six SM final states are produced per twin pion decay in this regime, the mean photon energy will scale down by a factor of three, but the multiplicity will increase by only slightly more than a factor of two. Moreover, any photon produced in a multi-body twin pion decay will have an energy of at most $m_{\twin{\pi}^0}/2$ in the twin pion rest frame, and therefore the photons produced by such decays cannot be harder, decay-by-decay, than for $\twin{\pi}^0 \to 2\gamma$ decays. The astrophysical $\gamma$-ray constraints typically loosen as a superlinear power law as the photon energy decreases, but only tighten linearly with multiplicity. Hence in the $3m_\pi < m_{\pi'} < 2m_K$ regime, for which the constraints in Sec.~\ref{sec:MC} do not apply, one expects the galactic center constraints in Fig.~\ref{fig:GCEcons} to be universally weaker over the entire parameter space, yielding larger allowed regions.

\section{Conclusions}
\label{sec:concl}
We have studied a twin WIMP (T-WIMP) scenario whose twin sector contains a mirror copy of the SM hadrons -- i.e. all three twin quark generations with three light quark flavors --  but no dark radiation -- i.e. no light twin leptons or photons. By comparison with existing twin Higgs frameworks, this scenario represents a relatively unexplored region of the twin Higgs theory space. 

Unlike various other twin Higgs frameworks, here the lightest twin degrees of freedom of this \emph{hadrosymmetric} twin Higgs model are twin pions -- pseudoscalars -- whose coupling to the SM sector via the Higgs portal is heavily suppressed. While the twin hadron sector can nevertheless be accessed at the LHC through the Higgs portal, any energy injected into the twin sector will just produce a shower of detector-stable, invisible particles. The only robust handles at the LHC, then, are evolving constraints on the invisible width of the Higgs.

From a cosmological and astrophysical point of view, however, the presence of light twin pions results in a much richer set of constraints and detection modes. To avoid a matter-dominated or overclosed universe post-BBN arising from metastable twin pions, the decay of the lightest hadron -- the $\pi'^0$ -- must be completed by the BBN epoch. For $m_{\pi'^0} < 3m_{\pi^0}$, this implies the presence of a new SM-twin mediation portal with a mass scale around a TeV or below. This new mediator can be probed at the intensity frontier or at the LHC in almost all of its viable parameter space. Despite the absence of dark radiation and LHC-accessible twin hadron decays, we nevertheless find a remarkable experimental complementarity between the cosmological bounds, the LHC and the intensity frontier.

In the Dirac and Majorana DM scenarios we consider, the dark matter freezes out through twin-electroweak mediated annihilation into twin quarks.  Annihilations of T-WIMPs then produce dark showers of twin hadrons which are either stable -- such as the $\twin{\pi}^\pm$ and twin proton -- or decay to SM photons -- the $\twin{\pi}^0$. Apart from Higgs invisible width bounds, these DM annihilations at the decoupling epoch are competitively constrained by CMB reionization bounds.  Such DM annihilations in the galactic center, however, naturally evade all astrophysical cosmic-ray constraints on antiproton or position production and proceed at a rate commensurate with a thermal relic cross-section, that in turn corresponds to a photon flux near the current sensitivities of \emph{Fermi}-LAT. Simulations of this dark showering process reveal that for the Dirac DM scenario, regions of parameter space that produce the observed DM abundances, as determined within the twin Higgs framework, and as preferred by naturalness and Higgs coupling constraints, are precisely those regions that successfully reproduce the claimed galactic center $\gamma$-ray excess in the \emph{Fermi}-LAT data. Conversely, current galactic center data, when applied as an upper bound on the total $\gamma$-ray flux from the galactic center, only partially constrains the available parameter space.

In this work we have mostly considered scenarios that do not admit a significant leptonic decay mode, $\pi'^0\rightarrow \mu^+\mu^-$. Should this decay channel be available, the required SM-twin portal can easily be out of reach of any near future experiment. However, the astrophysical bounds or reach for DM annihilations in the galactic center into high multiplicity soft leptonic final states are currently unknown. This may be explored in the future by conducting dedicated dark showering and galactic propagation simulations for this soft leptonic scenario.

\section*{Acknowledgments}
The authors thank Zackaria Chacko, David Curtin, Nathaniel Craig, Marco Farina, Andrey Katz, Eric Kuflik, John March-Russell, Michele Papucci, Tracy Slatyer, Matt Strassler, Wei Xue and Yue Zhao for helpful discussions. MF, SK and YT are grateful to the Galileo Galilei Institute for Theoretical Physics for hospitality during the preparation of part of this work. This work was prepared in part at the Aspen Center for Physics, which is supported by National Science Foundation (NSF) grant PHY-1066293. This work was supported in part by: the U.S. Department of Energy (DOE) under contract DE-SC0011640 (MF); the Office of High Energy Physics of the DOE under contract DE-AC02-05CH11231 (SK); the NSF under grant No. PHY-1002399 (DR); and the NSF under grant No. PHY-1315155 and the Maryland Center for Fundamental Physics (YT).

%

\end{document}